\documentclass[3p,times]{elsarticle}
\journal{}
\makeatletter
\def\ps@pprintTitle{%
 \let\@oddhead\@empty
 \let\@evenhead\@empty
 \def\@oddfoot{}%
 \let\@evenfoot\@oddfoot}
\makeatother
\usepackage{amsmath,amssymb}
\usepackage{tikz}
\usepackage{listings}
\usepackage{color}
\newcommand{\us}{\char`_}
\newcommand{\mc}[1]{\multicolumn{1}{c}{#1}}
\usepackage{colortbl}
\usepackage{algorithm}
\usepackage{algpseudocode,setspace}
\usepackage{hyperref}
\hypersetup{colorlinks=true,linkcolor=blue,filecolor=magenta,urlcolor=cyan}
\usepackage{xspace}
\newcommand{\nofft}{\emph{3D~MG}\xspace}
\newcommand{\yefft}{\emph{2D~MG~w/~FFT~(P$\leftrightarrow$S)}\xspace}
\newcommand{\yrfft}{\emph{3D~MG~w/~FFT}\xspace}
\newcommand{\yofft}{\emph{2D~MG~w/~FFT}\xspace}
\begin{document}

\begin{frontmatter}
  \title{A FFT-accelerated multi-block finite-difference solver for massively parallel simulations of incompressible flows\tnoteref{link}}
  \tnotetext[link]{Source code open and available under the terms of a MIT License on \href{https://github.com/p-costa/SNaC}{\nolinkurl{github.com/p-costa/SNaC}}.}
  \author[costa]{Pedro Costa}
  \ead{pcosta@hi.is}
  \address[costa]{Faculty of Industrial Engineering, Mechanical Engineering and Computer Science, University of Iceland, Hjardarhagi 2-6, 107 Reykjavik, Iceland}
  \begin{abstract}
    We present a multi-block finite-difference solver for massively parallel Direct Numerical Simulations (DNS) of incompressible flows. The algorithm combines the versatility of a multi-block solver with the method of eigenfunctions expansions, to speedup the solution of the pressure Poisson equation. This is achieved by employing FFT-based transforms along one homogeneous direction, which effectively reduce the problem complexity at a low cost. These FFT-based expansions are implemented in a framework that unifies all valid combinations of boundary conditions for this type of method. Subsequently, a geometric multigrid solver is employed to solve the reduced Poisson equation in a multi-block geometry. Particular care was taken here, to guarantee the parallel performance of the multigrid solver when solving the reduced linear systems equations. We have validated the overall numerical algorithm and assessed its performance in a massively parallel setting. The results show that $2$- to $8$-fold reductions in computational cost may be easily achieved when exploiting FFT acceleration for the solution of the Poisson equation. The solver, \emph{SNaC}, has been made freely available and open-source under the terms of an MIT license.
  \end{abstract}
\end{frontmatter}
%
%
\section{Introduction}\label{sec:intro}
Incompressible fluid flows abound in the nature and industry. From the nanoliter scales of the flow through capillary blood vessels, to the atmosphere dynamics at the planetary scale, there is a kaleidoscope of important phenomena with fluid dynamics in the leading role. Moreover, most fluid flows beyond the centimeter scale are in the turbulent state, exhibiting complex three-dimensional, chaotic dynamics that span a vast spectrum of scales. Indeed, this complexity has challenged generations of physicists and engineers to bridge the gap between our limited understanding of turbulent flows, and their prevalent nature. One of the main challenges stems from the nature of the Navier-Stokes equations governing fluid flows, which are unsteady, non-local, and highly non-linear, making its analysis extremely difficult.\par
Fortunately, the continuous developments of efficient numerical methods, together with the ever-increasing computing power \cite{top500}, enabled a paradigm-changing tool in fluid dynamics research: the Direct Numerical Simulations (DNS) of the Navier-Stokes equations. A DNS resolves all the scales of a fluid flow, providing a unique three-dimensional and time-resolved insight into their dynamics. Tremendous developments have followed the first DNS of homogeneous isotropic turbulence by \citet{Orszag-and-Patterson-PRL-1972} in 1972, being now possible to simulate canonical flows with trillions of spatial degrees of freedom \cite{Ishihara-et-al-ARFM-2009,Pirozzoli-et-al-JFM-2021}.\par
Finite-difference methods have been widely used in DNS of incompressible turbulent flows, particularly second-order, explicit finite-difference methods, following the seminal works of \citet{Kim-and-Moin-JCP-1985,Verzicco-and-Orlandi-JCP-1996}. Being typically very efficient, these methods can reproduce important observables of canonical turbulent flows with high fidelity \cite{Vreman-and-Kuerten-PoF-2014,Moin-and-Verzicco-EJMBF-2016}, while remaining versatile in terms of the types of geometries, computational grids, boundary conditions, and the incorporation of more complex phenomena. In fact, several works have shown that high-quality data obtained from second-order, explicit finite-difference methods are not necessarily of inferior quality compared to that obtained from very high-order calculations (see, e.g., \cite{Moin-and-Verzicco-EJMBF-2016}), as long as proper (higher) resolution is secured. Indeed, combined with immersed boundary methods to simulate the flow over complex geometries \cite{Fadlun-et-al-JCP-2000,Breugem-and-Boersma-PoF-2005,Uhlmann-JCP-2005}, interface-tracking/-capturing methods for multi-fluid flows \cite{Tryggvason-et-al-2011,Aniszewski-et-al-CPC-2021}, or to simulate canonical flows at very high Reynolds numbers \cite{Pirozzoli-et-al-JFM-2021}, this class of finite-difference methods has been playing a major role in DNS.\par
The incompressible Navier-Stokes equations have a highly non-local nature, due to the need to couple a constraint of zero velocity divergence -- mass conservation -- to the momentum transport equation. This typically involves a solution of a Poisson equation for a pressure field, which is used to project the velocity field into a divergence-free space \cite{Chorin-MC-1968}. The Poisson equation encapsulates the main challenge of solving the incompressible Navier-Stokes equations in a massively parallel framework -- any disturbance in the system is propagated instantly and everywhere by the pressure. Indeed, the Poisson solver is typically the most expensive and elaborate part of an incompressible DNS solver.\par
Geometric multigrid methods have been proving to be efficient in solving the second-order finite-difference Poisson equation \cite{Wesseling-2009,Golub-book}. These methods exhibit excellent scaling properties, allow for non-uniform grids, and are versatile in the boundary conditions that can be accommodated. In relatively simple domains, however, very efficient direct solvers can be used instead, e.g.\ by exploiting the method of eigenfunctions expansions \cite{Swarztrauber-SR-1977,Schumann-and-Sweet-JCP-1988}. This method uses Fourier-based expansions which reduce the number of diagonals of the linear system in two domain directions, resulting in a simple tridiagonal system that can be efficiently solved with Gauss elimination \cite{Swarztrauber-and-Sweet-JCAM-1989}.
Thanks to the continuous improvements of frameworks for the development of parallel algorithms, this approach has regained popularity and has been employed in numerous recent studies \cite{Costa-CAMWA-2018}. Indeed, this method has allowed for breakthroughs in e.g.\ DNS of single-phase canonical turbulent flows \cite{Ostilla-et-al-JFM-2016,Pirozzoli-et-al-JFM-2021}, in complex geometries by using immersed boundary methods \cite{Breugem-and-Boersma-PoF-2005}, and in multi-phase flows \cite{Costa-et-al-PRL-2016,Dodd-and-Ferrante-JFM-2016,Cifani-et-al-CF-2018}, with at least two open-source DNS codes, \emph{AFiD} \cite{Van-der-Poel-et-al-CF-2015} and \emph{CaNS} \cite{Costa-CAMWA-2018}, leveraging this approach. Despite most works in the literature only exploiting the method of eigenfunctions expansions along periodic directions \cite{Borrell-et-al-JCP-2011}, these Fourier-based expansions may be actually employed for many different combinations of boundary conditions \cite{Schumann-and-Sweet-JCP-1988}.\par
To our best knowledge, finite-difference numerical algorithms reported in the literature using FFT-based finite-difference solvers are restricted to very simple geometries such as a rectangular box \cite{Laizet-and-Li-IJNMF-2011,Costa-CAMWA-2018} or cylindrical/spherical domains \cite{Van-der-Poel-et-al-CF-2015,Santelli-et-al-JCP-2021}, which may be extended to handle more complex geometries using immersed boundary methods \cite{Mittal-and-Iaccarino-ARFM-2005}. Despite their proven fidelity to treat complex geometries efficiently, single-box solvers with immersed boundary methods may not be optimal for cases where a substantial portion of the computational domain is masked by the immersed solid volume (e.g., a narrow T-junction type of geometry), due to a large number of superfluous calculations outside the physical domain. This type of geometries may be better suited for a solver that can be partitioned into multiple boxes, or \emph{blocks}, to solve the Navier-Stokes equations only in the relevant physical domain.\par
The present work aims precisely to relax the restriction of current high-fidelity finite-difference DNS solvers, while retaining the versatility and efficiency of FFT-based synthesis of the Poisson equation. To this goal, we present an efficient multi-block Navier-Stokes solver for massively parallel simulations of fluid flows. The solver may leverage the method of eigenfunctions expansions to solve the Poisson equation along one homogeneous ``\emph{extruded}'' direction, decoupling the systems of equations in that direction, and employs highly efficient geometric multigrid solvers \cite{Falgout-et-al-2002} for the reduced systems of equations. Similarly to the DNS code \emph{CaNS}, the FFT-based expansion is implemented so as to cover all valid combinations of boundary conditions. The resulting tool, \emph{SNaC}, has been made freely available and open-source.\par
We present the design and implementation of the algorithm in a massively parallel framework, with adaptations to leverage the \emph{hypre} library of multigrid solvers to solve the reduced Poisson equation after FFT-based synthesis. The results illustrate the high efficiency and versatility of this approach in different systems, resulting in up to an $8$-fold speedup of the numerical calculation. Hence, in the same spirit as efficient single-block codes such as \emph{CaNS} and \emph{AFiD}, \emph{SNaC} serves as a good base multi-block DNS solver, on top of which extensions to handle more complex physics such as two-phase flows or irregular geometries can also be implemented.\par
Next, in \S\ref{sec:numerics}, we will describe the governing equations and numerical method. Then \
\S\ref{sec:implm} presents our general implementation strategy, and the approach to enable simulations in a massively parallel setting. We will then present in \S\ref{sec:results} the validation of the numerical algorithm, and assess its performance. Finally, \S\ref{sec:conclusions} provides a summary and future perspectives.
\section{Governing Equations and Numerical Method}\label{sec:numerics}
The numerical algorithm solves the incompressible Navier-Stokes equations for a fluid with unit density $\rho=1$ and kinematic viscosity $\nu$,
\begin{align}
  \boldsymbol{\nabla} \cdot \mathbf{u} = 0 \mathrm{,} \label{eqn:cont} \\
  \frac{\partial \mathbf{u}}{\partial{t}} + \nabla\cdot(\mathbf{u} \otimes\mathbf{u}) = - \nabla P + \nu\nabla^2{\mathbf{u}}\mathrm{,}\label{eqn:mom}
\end{align}
with $\mathbf{u}$ and $P$ being the fluid velocity vector and pressure.\par
These equations are discretized using a second-order finite-difference/finite-volume method on a structured Cartesian grid, with staggered flow variables \cite{Harlow-and-Welch-PoF-1965} to avoid odd-even decoupling phenomena and preserve energy at the discrete level (in the inviscid limit) \cite{Verstappen-and-Veldman-JCP-2003}. The grid spacing may vary along any direction that does not exploit an FFT-based synthesis of the Poisson equation (described later in this section). The equations are integrated in time using a low-storage three-step Runge-Kutta scheme (RK3) in a standard fractional-step method \cite{Chorin-MC-1968,Kim-and-Moin-JCP-1985,Rai-and-Moin-JCP-1991}. The time advancement an is fully explicit, and reads at each substep $k$ ($k=1,2,3;\, k=1$ corresponds to a time level $n$ and $k=3$ to $n+1$):
\begin{align}
  \mathbf{u}^* = \mathbf{u}^k + \Delta t\left(\alpha_k\left(\mathcal{A}\mathbf{u}^{k}+\nu\mathcal{L}\mathbf{u}^{k}\right) + \beta_k\left(\mathcal{A}\mathbf{u}^{k-1} + \nu\mathcal{L}\mathbf{u}^{k-1}\right) -      \gamma_k\mathcal{G} P^{k-1/2}\right)\mathrm{,} \label{eqn:up} \\
  \mathcal{L}\Phi = \frac{\mathcal{D}\mathbf{u}^*}{\gamma_k\Delta t}\mathrm{,}\label{eqn:poi_ns}                                                                                                                                                                                  \\
  \mathbf{u}^{k+1} = \mathbf{u}^* - \gamma_k\Delta t \mathcal{G}\Phi\mathrm{,}                                                                                                                                                                                                        \\
  P^{k+1/2} = P^{k-1/2} + \Phi\mathrm{,}
\end{align}
where $\mathcal{A}$, $\mathcal{L}$, $\mathcal{G}$, and $\mathcal{D}$ denote the discrete advection, Laplacian, gradient, and divergence operators; $\mathbf{u}^*$ is the prediction velocity and $\Phi$ the correction pressure. The RK3 coefficients are given by $\alpha=\lbrace{8/15,5/12,3/4\rbrace}$, $\beta=\lbrace{0,-17/60,-5/12\rbrace}$, and $\gamma=\alpha+\beta$. A sufficient criterion for a stable temporal integration is given in \cite{Wesseling-2009}:
\begin{equation}
  \Delta t \le \min\left(\frac{1.65\Delta \ell_{\min}^2}{\nu},\frac{\sqrt{3}\Delta \ell_{\min}}{\max {||\mathbf{u}||_1}}\right)\mathrm{,}
\end{equation}
with $||\mathbf{u}||_1$ the $\ell_1$-norm of $\mathbf{u}$, and $\Delta \ell_{\min}$ the smallest grid spacing. Optionally, the temporal integration of the diffusion term may be treated implicitly. To achieve that, we directly solve three additional Helmholtz equations using the same numerical method that is used for the Poisson equation, even though a more efficient alternating diagonal implicit (ADI) approach could also be employed \cite{Peaceman-and-Rachford-1955,Kim-and-Moin-JCP-1985}.
\subsection*{Poisson Solver}
One essential feature of the present method concerns the solution of the Poisson equation for the correction pressure $\Phi$. The equation at grid point $i,j,k$ reads, assuming constant grid spacing in each direction for simplicity,
\begin{align}
   & (\Phi_{i-1,j  ,k  }-2\Phi_{i,j,k}+\Phi_{i+1,j  ,k  })/\Delta x_1^{2} + \nonumber                          \\
   & (\Phi_{i  ,j-1,k  }-2\Phi_{i,j,k}+\Phi_{i  ,j+1,k  })/\Delta x_2^{2} + \nonumber                          \\
   & (\Phi_{i  ,j  ,k-1}-2\Phi_{i,j,k}+\Phi_{i  ,j  ,k+1})/\Delta x_3^{2} = f_{i,j,k}\mathrm{,}\label{eqn:poi}
\end{align}
which corresponds to a linear system represented by a Poisson matrix with $7$ non-zero diagonals; $\Delta x_l$ denotes the grid spacing in direction $x_l$ ($l = 1$, $2$, or $3$). Here we exploit the method of eigenfunctions expansions to reduce the complexity of the Poisson equation by decoupling it along one direction, say $x_2$. To achieve this \emph{Fourier synthesis} \cite{Schumann-and-Sweet-JCP-1988}, a Fourier-based discrete expansion operator, $\mathcal{F}_{x_l}$, is employed to Eq.~\eqref{eqn:poi}, resulting in the following Helmholtz equation
\begin{align}
  - \left(\frac{2}{\Delta x_1^2} + \frac{2}{\Delta x_3^2} - \frac{\lambda_{j}}{\Delta x_2^2}\right) \hat{\Phi}_{i,j,k}
  + \frac{\hat{\Phi}_{i-1,j,k}+\hat{\Phi}_{i+1,j,k}}{\Delta x_1^{2}} + \frac{\hat{\Phi}_{i,j,k-1}+\hat{\Phi}_{i,j,k+1}}{\Delta x_3^{2}} = \hat{f}_{i,j,k}\mathrm{,} \label{eqn:poi_reduced}
\end{align}
where $\hat{\square} = \mathcal{F}_{x_l}(\square)$, denotes the Fourier-based discrete transform along direction $x_l$, and $\lambda_q$ is the eigenvalue associated with the wavenumber $\kappa_q$ ($q=0,\dots,n_l-1$ with $n_l$ the number of grid points along $x_l$). The eigenfunction expansion $\mathcal{F}_{x_l}$ and eigenvalues $\lambda_q$ depend on the boundary conditions at each end of the expansion direction, which have to be satisfied by the corresponding inverse operator $\mathcal{F}_{x_l}^{-1}$. For instance, $\mathcal{F}_{x_l}$ would be the discrete Fourier transform in case of periodic boundary conditions, or a discrete sine transform in case of Dirichlet boundary conditions at both ends. Indeed, various eigenfunction expansions and eigenvalues for different combinations of boundary conditions may be employed. The types of direct and inverse discrete transforms $\mathcal{F}_{x_l}$ and corresponding eigenvalues $\lambda_q$ for different combinations of pressure boundary conditions are listed in Table~\ref{tbl:operators}, and we refer to e.g.~\cite{Schumann-and-Sweet-JCP-1988,Fuka-AMC-2015,Costa-CAMWA-2018} for more details.
\begin{table}[hbt!]
  \centering
  \caption{Eigenvalues, and forward ($\mathcal{F}$) and backward ($\mathcal{F}^{-1}$, multiplied by the normalization factor $\theta$) transforms, to solve Eq.~\eqref{eqn:poi_reduced} for different combinations of staggered boundary conditions~\cite{Schumann-and-Sweet-JCP-1988}. The eigenvalues in Eq.~\eqref{eqn:poi_reduced} are given by $\lambda_q = 2\,(\cos(\kappa_q\Delta \ell)-1)$, $q=0,\dots,n-1$, with $n$ being the (even) number of grid points in the direction of synthesis, $\kappa_q$ a wavenumber, and $\Delta \ell = \ell/n$ the corresponding grid spacing for a domain with length $\ell$. The mathematical expressions for the different transforms can be found in, e.g., \cite{Costa-CAMWA-2018}. Here \texttt{(I)DFT} denotes the (inverse) discrete Fourier transform, and \texttt{DST}/\texttt{DCT} the different standard types of discrete sine/cosine transforms.}\label{tbl:operators}
  \begin{tabular}{ l l l l l l }
    \hline
    \hline
    \mc{Boundary Conditions}    & \mc{$\kappa_q/(2\pi/\ell)\,\,[\lambda_q = 2(\cos(\kappa_q\Delta \ell)-1)]$} & \mc{$\mathcal{F}$} & \mc{$ \theta\,\mathcal{F}^{-1}$} & \mc{$\theta$} \\
    \hline
    \hline
    \small Periodic             & $q$                                                                         & \texttt{DFT   }    & \texttt{IDFT   }                 & $~n$          \\
    \hline
    \small Neumann--Neumann     & $q/2$                                                                       & \texttt{DCT-II}    & \texttt{DCT-III}                 & $2n$          \\
    \hline
    \small Dirichlet--Dirichlet & $(q+1)/2$                                                                   & \texttt{DST-II}    & \texttt{DST-III}                 & $2n$          \\
    \hline
    \small Neumann--Dirichlet   & $(2q+1)/4$                                                                  & \texttt{DCT-IV}    & \texttt{DCT-IV }                 & $2n$          \\
    \hline
    \hline
  \end{tabular}
\end{table}\par
The advantage of the Fourier synthesis of Eq.~\eqref{eqn:poi} is that all discrete transforms presented in Table~\ref{tbl:operators} may exploit the FFT algorithm, resulting in a relatively low cost of $O(N\log n_2)$ operations, with $N$ the total number of grid points, and $n_2$ the number of grid points along $x_2$. Note, however, that the grid is required to be uniform in the direction of synthesis.\par
In simple rectangular boxes, it is beneficial to further simplify this equation by employing this Fourier synthesis in a second direction, say in $x_3$. With a total cost of $O(N\log n_2 n_3)$ operations \cite{Costa-CAMWA-2018}, the two reductions enable an efficient, \emph{direct} solution of the Poisson equation -- the problem is simplified to the solution of $n_2 n_3$ tridiagonal systems with $n_1$ unknowns ($O(n_1 n_2 n_3) = O(N)$ operations). This was the approach used in the DNS solver \emph{CaNS} \cite{Costa-CAMWA-2018}, and showed excellent performance. In a multi-block domain, instead, the geometry is expected to be more complex, with the number of grid points $n_1$ and $n_2$ varying among blocks. This makes a two-dimensional FFT-based synthesis impractical to implement in a distributed-memory framework. Even so, employing this synthesis in one direction to obtain Eq.~\eqref{eqn:poi_reduced} is often possible and desirable -- there are numerous interesting cases where a multi-block, two-dimensional configuration that is ``extruded'' along a third direction, such as a T-junction, a cross-slot, a square elbow type of geometry. These are precisely the type of geometries, homogeneous along one direction, which can benefit from the FFT-based acceleration of the Poisson equation in the present method.\par
It is interesting to note that the computational complexity of efficient iterative methods such as a geometric multigrid solver for Eq.~\eqref{eqn:poi} scales with $O(N)$, while a direct solution with Fourier synthesis in two directions scales less efficiently, $O(N\log n_2 n_3)$. Interestingly, so far, FFT-based direct solvers of a Poisson equation, with $N\sim 10^{9}-10^{10}$ have been reported to yield excellent performance $3-10$ times faster than well-established a geometric multigrid solvers (depending on the type of solver and desired tolerance; see, e.g., \cite{Ahmed-et-al-CF-2020}). While this trend is expected to reverse for sufficiently high values of $N$, the term $\log n_2 n_3 $ grows slowly, meaning current ambitious problem sizes may still be orders-of-magnitude too small for efficient iterative methods to overperform direct FFT-based solvers.\par
In the absence of Fourier synthesis, Eq.~\eqref{eqn:poi} is solved using the efficient parallel semicoarsening multigrid solver PFMG (which uses a point-wise smoother), or the more robust SMG solver (which uses a plane smoother) \cite{Falgout-et-al-2002}, available in the \emph{hypre} library. The PFMG solver estimates the best direction of semicoarsening by choosing the smallest grid spacing direction (or attempts to coarsen along $x_1$, then $x_2$, and then $x_3$ for equal values of smallest grid spacing along the different directions). Weighted Jacobi (used in the present work), or red/black Gauss-Seidel may be used for the smoother. On the other hand, the SMG solver coarsens along $x_3$, and smooths along $x_1-x_2$ planes. This planar smoothing uses a single 2D SMG cycle, which in turn coarsens along $x_2$, and uses $x_1$-line smoothing. See, e.g., \cite{Ashby-and-Falgout-NSE-1996,Baker-et-al-2012} for more details.\par
When Fourier synthesis is employed, the same solvers are used to solve the resulting decoupled two-dimensional Helmholtz equations. It is important to note that the magnitude of the diagonal elements of the matrix corresponding to each two-dimensional system, Eq.~\eqref{eqn:poi_reduced}, will vary according to $\lambda_j$ (recall Table~\ref{tbl:operators}). Hence, the iterative solution convergence is expected to vary among the two-dimensional systems \cite{Golub-book}, requiring a larger number of iterations for smaller values of $\lambda_j$.\par
For clarity, the steps undertaken to solve the Poisson equation in this case are described below, in Algorithm~\ref{alg:fast_poi_solver}, for a square box with dimensions $n_1\times n_2\times n_3$ and Fourier synthesis along $x_2$.
\begin{algorithm}[hbt!]
  \caption{Summary of the steps required for solving Eq.~\eqref{eqn:poi} in a $n_1\times n_2\times n_3$ box, using Fourier synthesis along $x_2$.\label{alg:fast_poi_solver}}
  \begin{algorithmic}
    \State \textbf{do} {$i=1$ to $n_1$ and $k=1$ to $n_3$} \Comment{$n_1 n_3\cdot O(n_2 \log n_2)$ operations}
    \State \quad forward FFT-based transform along $x_2$ of right-hand-side of Eq.~\eqref{eqn:poi}: $\hat{f}_{i,1 \dots n_2,k} = \mathcal{F}_{x_2}(f_{i,1 \dots n_2,k})$
    \State \textbf{end do}
    \State \textbf{do} $j=1$ to $n_2$ \Comment{$n_2\cdot O(n_1 n_3)$ operations.}
    \State \quad solve Eq.~\eqref{eqn:poi_reduced} using a geometric multigrid solver to obtain $\hat{\Phi}_{1 \dots n_1,j,1 \dots n_3}$
    \State \textbf{end do}
    \State \textbf{do} $i=1$ to $n_1$ and $k=1$ to $n_3$ \Comment{$n_1 n_3 \cdot O(n_2\log n_2)$ operations}
    \State \quad backward FFT-based transform along $x_2$ of the previous solution: $\Phi_{i,1 \dots n_2,k} = \mathcal{F}_{x_2}^{-1}(\hat{\Phi}_{i,1 \dots n_2,k})$
    \State \textbf{end do}
  \end{algorithmic}
\end{algorithm}\par
Important implementation details for solving these equations in a massively parallel paradigm will follow next.
\section{Implementation Strategy}\label{sec:implm}
The numerical tool has been implemented in modern Fortran, and extended with MPI/OpenMP for distributed- and shared-memory parallelization. The OpenMP extension serves to guide future porting efforts to heterogeneous (e.g., many-GPU) systems, which may exploit directive-based approaches for thread-level parallelism; its performance will not be discussed here.\par
\subsection{Computational setup}\label{sec:comput_setup}
\begin{figure}[hbt!]
  \centering
  \includegraphics[width=0.55\columnwidth]{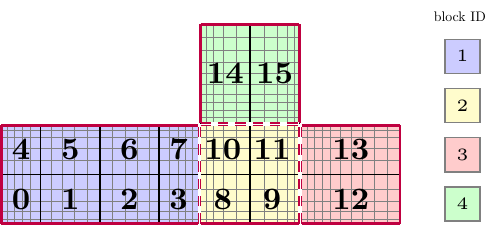}
  \caption{Illustration of a valid multi-block setup in two dimensions. The red solid lines denote physical boundaries, while the dashed lines denote internal boundaries. The four blocks (depicted in different colors) are partitioned into several computational subdomains (depicted by the different numbers). The non-uniform grid was obtained using hyperbolic tangent based mapping functions; see~\cite{Orlandi-book}. The computational parameters used to generate this grid are reported in Table~\ref{tbl:illustration_block}, and described in its caption.}\label{fig:illustration_block}
\end{figure}

The problem is set by two kinds of computational parameters -- \emph{global} and \emph{block-specific}. Global parameters are those common to all blocks, such as physical properties and reference scales, time step control, simulation stopping criteria and I/O frequency; block-specific parameters set, for each block, the geometry and computational mesh, the boundary conditions (including inter-block connectivity), and the three-dimensional block domain partitioning into different computational subdomains, each assigned to an MPI process. These parameters have to be set such that the grid along the boundaries of connected blocks is congruent, so the whole computational domain is discretized on a structured grid. Moreover, the partitioning into different computational subdomains is conditioned to the following rules:
\begin{itemize}
  \item[--] blocks can be decomposed in the three domain directions, and each MPI process is assigned \emph{exclusively} to one of the corresponding computational subdomains. Consequently, each block needs to be assigned to at least one MPI process;
  \item[--] each side of a computational subdomain is either a physical boundary, or is connected to a single neighboring subdomain;
  \item[--] if FFT-based synthesis of the Poisson equation is used, the computational subdomains cannot be decomposed along the direction of synthesis (i.e., a pencil-like domain decomposition is required).
\end{itemize}
Fig.~\ref{fig:illustration_block} presents an example of a valid computational setup in two dimensions, where the geometry is partitioned into $4$ blocks and a total of $16$ computational subdomains. As the figure illustrates, MPI ranks are grouped consecutively within each block, with row-major ordering. Those partitions are set by a block-specific input parameter dictating the number of subdivisions in each direction. The partitioning is then performed so as to distribute as evenly as possible the block grid cells among the different subdomains, along each direction. More specifically, for $n_l$ points partitioned into $m_l$ MPI tasks along direction $x_l$, the first $n_l \% m_l$ tasks will manage $\lfloor n_l/m_l\rfloor+1$ grid points, and the remaining tasks will manage $\lfloor n_l/m_l \rfloor$ points.\par
In practice, the blocks are defined as illustrated in Table~\ref{tbl:illustration_block} (corresponding to the setup in Fig.~\ref{fig:illustration_block}). First, the coordinates of the lower and uppermost corners of each block (\texttt{lo} and \texttt{hi}) are defined in index space, i.e., in a coordinate system with arbitrary origin and uniform spacing equal to $1$, such that the number of grid points in each direction is equal to \texttt{hi-lo+1}. Then the physical coordinates of each of the corners are defined by parameters \texttt{lmin} and \texttt{lmax}, and a mapping function of choice is used to determine the coordinates of the grid points in the physical coordinate system (e.g., to achieve a grid clustering bias). In Table~\ref{tbl:illustration_block}, \texttt{gr.~type} defines the mapping function type, and \texttt{gr.~factor} is a parameter dictating the degree of clustering. Finally, \texttt{dims} sets the number of partitions of the block, in each direction. The caption of Table~\ref{tbl:illustration_block} explains in more detail how those parameters result in the configuration of Fig.~\ref{fig:illustration_block}. We should note that a two-dimensional system is naturally obtained from a three-dimensional setup using two grid cells and a small domain length along one direction, with a two-dimensional initial condition.
\begin{table}[hbt!]
  \centering
  \caption{Block-specific parameters for the two-dimensional configuration in Fig.~\ref{fig:illustration_block}. \texttt{lo} and \texttt{hi} are the index space coordinates of the lower and uppermost corners of each block; \texttt{lmin} and \texttt{lmax} the physical coordinates of these corners; \texttt{gr.~type} denotes the choice of grid mapping functions, which have to be congruent among blocks (among the family of functions implemented in \emph{SNaC}, we illustrate here hyperbolic tangent clustering at two ends (\texttt{0}), or just at the lower/upper end \texttt{-1/+1}); \texttt{gr.~factor} is the grid stretching parameter of the mapping function, with \texttt{0.0} corresponding to a uniform mapping (see \cite{Orlandi-book}); \texttt{dims} dictates the MPI partitioning along each direction.}\label{tbl:illustration_block}
  \small
  \begin{tabular}{c c c c c c c c}
    \hline
    \hline
    block ID                             & \texttt{lo}         & \texttt{hi}        & \texttt{lmin}       & \texttt{lmax}        & \texttt{gr.~type}   & \texttt{gr.~factor}  & \texttt{dims}    \\
    \hline
    \hline
    \colorbox{blue!20!white}{\texttt1}   & \texttt{[~1,~1,~1]} & \texttt{[20,10,2]} & \texttt{[0.,0.,0.]} & \texttt{[2.,1.,0.1]} & \texttt{[ 0, 0, 0]} & \texttt{[2.5,1.,0.]} & \texttt{[4,2,1]} \\
    \colorbox{yellow!20!white}{\texttt2} & \texttt{[21,~1,~1]} & \texttt{[30,10,2]} & \texttt{[2.,0.,0.]} & \texttt{[3.,1.,0.1]} & \texttt{[-1,-1, 0]} & \texttt{[1.5,1.,0.]} & \texttt{[2,2,1]} \\
    \colorbox{red!20!white}{\texttt3}    & \texttt{[31,~1,~1]} & \texttt{[40,10,2]} & \texttt{[3.,0.,0.]} & \texttt{[4.,1.,0.1]} & \texttt{[ 1, 1, 0]} & \texttt{[1.5,1.,0.]} & \texttt{[1,2,1]} \\
    \colorbox{green!20!white}{\texttt4}  & \texttt{[21,11,~1]} & \texttt{[30,20,2]} & \texttt{[2.,1.,0.]} & \texttt{[3.,2.,0.1]} & \texttt{[ 0, 0, 0]} & \texttt{[2.5,1.,0.]} & \texttt{[2,1,1]} \\
    \hline
    \hline
  \end{tabular}
\end{table}\par
Finally, physical and block-block boundary conditions need also to be specified. Three kinds of boundary conditions may be set for the velocity and pressure -- Dirichlet, Neumann, or block-block connectivity, with periodic boundary conditions being naturally set by a cyclic sequence of connectivity conditions along one direction. Naturally, the velocity and pressure boundary conditions need to be consistent, so that the pressure projection step at the boundary yields the expected normal velocity component (e.g., a prescribed velocity requires a zero normal gradient of $\Phi$).
\subsection{Overview of the parallel implementation strategy}\label{sec:par_strategy}
The following steps are performed to set up the calculation in a distributed-memory framework:
\begin{enumerate}
  \item \textbf{Assign MPI tasks to the computational subdomains:} for each block, subsets of the total number of MPI processes (hereafter denoted \verb|comm_world|) are assigned to each computational subdomain, and the corresponding local grid spacing and extents are determined as illustrated in the previous section;
  \item \textbf{Determine neighboring MPI tasks:} for each computational subdomain, the task IDs of the six neighboring subdomains (i.e., $2$ per domain direction) are determined and stored (with \verb|MPI_PROC_NULL| tagging a non-cyclic physical boundary);
  \item \textbf{Describe data structures for boundary data exchange:} data structures for ghost cells communication among neighboring tasks are created (\verb|MPI_Type_vector| describing the boundary data layout), as well as a communicator \verb|comm_block| grouping the tasks per block, to be used for post-processing and I/O.
\end{enumerate}
Once these initialization steps are performed and the neighbors of each MPI process determined, \emph{the algorithm becomes agnostic of the disposition of blocks} -- communication of ghost cell data between neighboring computational subdomains (so-called halo exchange) may be performed with, e.g., a \verb|MPI_Sendrecv| call, without discerning internal and external block boundaries.\par
Finally, MPI-I/O is used to write field data into a single binary file per block, which is accompanied by a file logging the saved data information. This allows visualizing field data as a time series using a simple \emph{XDMF} metadata file \cite{xdmf}. For all cases assessed here, the MPI-I/O implementation performed well, with a time for checkpointing comparable to that of one calculation time step.

\subsection{Massively parallel Poisson solver}
The different solution strategies for solving Eq.~\eqref{eqn:poi} on a multi-block geometry are described below\footnote{The implementation is actually more general, solving a Helmholtz equation on non-uniform structured Cartesian grids, with staggered or non-staggered boundary conditions.}. A common denominator in these approaches is the efficient and well-established \emph{hypre} library of high-performance multigrid solvers. Indeed, the library's Structured-Grid-System (\emph{Struct}) conceptual interface for structured-grid applications enabled a versatile implementation, however with excellent performance. It should be noted that the implementation in \emph{SNaC} allows for flexibility in the choice of the direction of FFT-based synthesis (or no synthesis at all) by employing (\verb|cpp|) source pre-processing.
\subsubsection{Geometric multigrid solver without FFT-based synthesis}\label{sec:plain_3d_mg}
Solving Eq.~\eqref{eqn:poi} without FFT-based synthesis is a canonical use case of the \emph{hypre}'s \emph{Struct} interface. In a nutshell, the interface defines a distributed coefficient matrix by passing to the library:
\begin{enumerate}
  \item the MPI communicator where the calculation is to be performed (here, \verb|comm_world|);
  \item the extent of each computational subdomain in index space (same convention as parameters \texttt{lo} and \texttt{hi} in Table~\ref{tbl:illustration_block});
  \item information about the finite-difference stencil associated with the system;
  \item the $7$ non-zero elements of the coefficient matrix (one per stencil entry), for each grid point within the computational subdomain.
\end{enumerate}
Subsequently, the setup of the right-hand side and initial guess vectors, and the setup of the geometric multigrid solver are straightforward.
These initialization steps are performed once at the beginning of the calculation\footnote{If implicit temporal discretization of the diffusion term is used, not discussed here, the coefficient matrix diagonal needs to be modified at every RK3 substep, which is possible using \emph{hypre}'s \texttt{HYPRE{\us}StructMatrixAddToBoxValues}.}; the Poisson equation is then solved every RK3 substep using the latest solution as the initial guess.
\subsubsection{FFT-accelerated solution of the Poisson equation}\label{sec:mg_w_fft}
The FFT-accelerated solution of the Poisson equation described in Algorithm~\ref{alg:fast_poi_solver} can be employed as long as the domain has one homogeneous ``extruded'' direction with constant grid spacing. We adopted the implementation of FFT-based synthesis in \emph{CaNS} \cite{Costa-CAMWA-2018}, which uses the \emph{guru} interface of the \emph{FFTW} library~\cite{FFTW3}. This approach computes all types of fast discrete transforms in Table~\ref{tbl:operators} efficiently, in place, and with the same syntax, just by evoking the right transform type and considering the different scaling factors.\par
\begin{figure}[hbt!]
  \centering
  \begin{tikzpicture}
    \node[draw=none,fill=none] at (0,0) {\includegraphics[width=0.55\columnwidth]{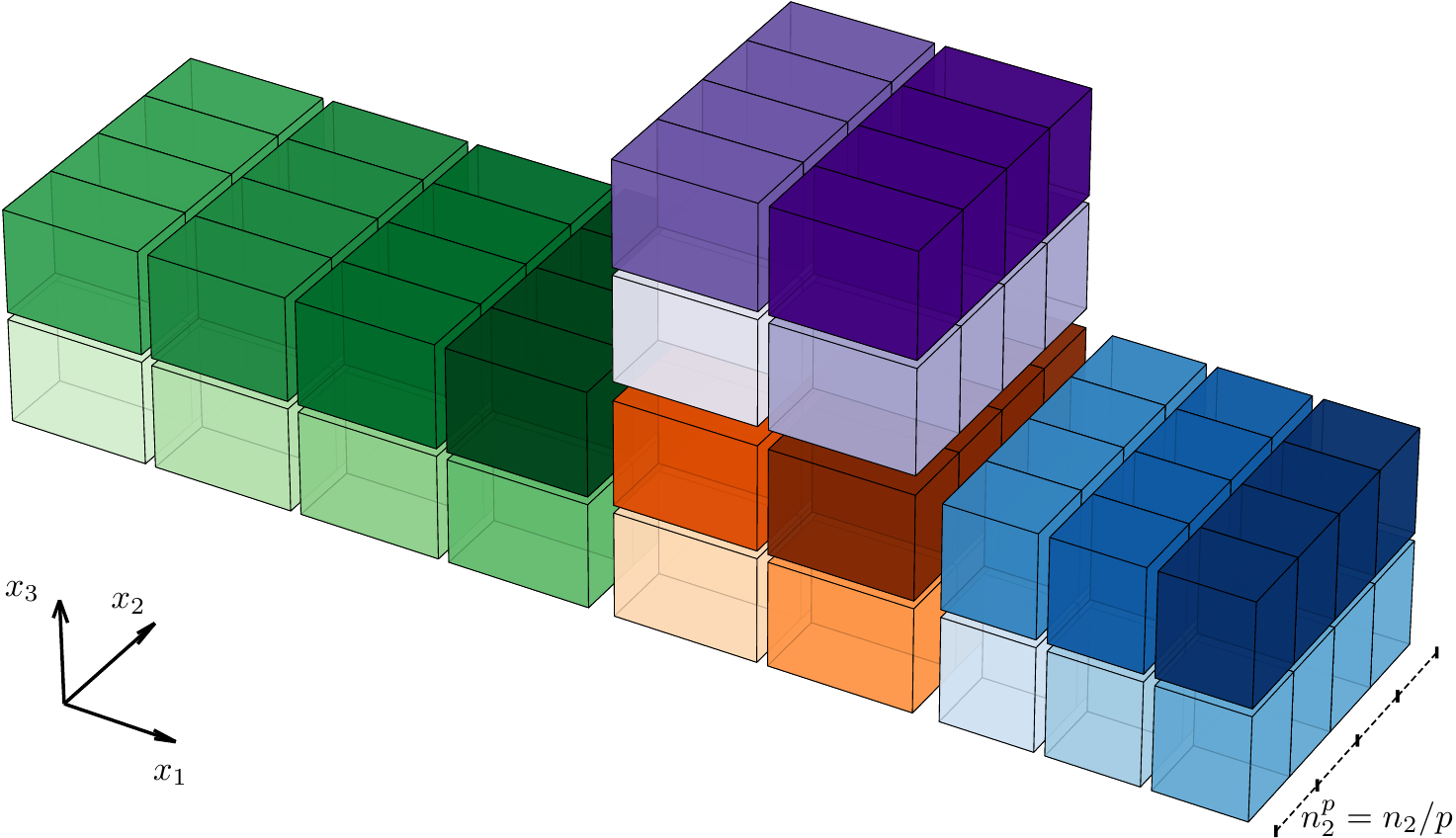}};
    \node at (-0.50,-2.5) {\scriptsize grid cells for \emph{pencil} subdomain $m$: $\left[n_1^m\times n_2\times n_3^m\right]$~~};
  \end{tikzpicture}
  \caption{Illustration of the domain decompositions to solve the FFT-accelerated Poisson equation, assuming $x_2$ as the FFT synthesis direction. Different colors distinguish the tasks in different blocks, while different lightness marks the MPI tasks within a block. After performing the FFT-based synthesis, to solve the resulting $n_2$ independent 2D linear systems, the pencils are partitioned by a factor $p$ (here $p=4$), and a 3D system decoupled along $x_2$ is defined for each chunk.}\label{fig:sliced_pencils}
\end{figure}

As illustrated in Algorithm~\ref{alg:fast_poi_solver}, the first step is performing one-dimensional FFT-based transforms along the homogeneous direction, here taken as $x_2$. To achieve this in a distributed memory paradigm, the domain is not decomposed along $x_2$, as illustrated in Fig.~\ref{fig:sliced_pencils}. In this \emph{pencil} decomposition, each computational subdomain $m$ has a size $\left[n_1^{m},n_2,n_3^{m}\right]$.\par
After employing the one-dimensional FFT-based transforms, $n_2$ \emph{decoupled} 2D systems will be solved using the geometric multigrid method (recall Eq.~\eqref{eqn:poi_reduced}), with each system set analogously to the 3D system described above in \S\ref{sec:plain_3d_mg}. Three approaches were considered:

\begin{itemize}
  \item[--] \emph{\textbf{The naive approach.}} Using the pencil decomposition, these 2D linear systems can be solved consecutively, parallelized over \verb|comm_world|, i.e., solving for $\hat{\Phi}_{1 \dots n_1^m,j,1 \dots n_3^m}$, from $j=1$ to $n_2$. However, as we will see, solving such small linear systems in a massively parallel setting will result in a significant communication overhead, with all tasks synchronizing between each solve. Moreover, it is not yet possible to set explicitly a batch of systems to be solved collectively using the \emph{hypre} library.\par
  \item[--] \emph{\textbf{The sliced pencils approach.}} To circumvent this issue, we define batches of 2D systems as small 3D problems -- 3D linear systems are set as previously described, but decoupled along $x_2$ by setting the stencil coefficients in this direction to zero. Care should be taken here, because the number of iterations to solve each 2D system varies along $x_2$, due to the eigenvalue $\lambda_{j}$ in the diagonal of each system (recall Eq.~\eqref{eqn:poi_reduced} and Table~\ref{tbl:operators}). If, for instance, a single distributed 3D matrix encapsulating the entire pencil subdomain with size $\left[n_1^{m},n_2,n_3^{m}\right]$ is considered, much unnecessary work will be performed in the 3D problem, to match the maximum number of iterations of the slowest-converging 2D system. Hence, to cover the problem inhomogeneity along $x_2$, the pencil subdomains are sliced into $p$ chunks, hence with a size $\left[n_1^m,n_2^p,n_3^m\right]$ with $n_2^p=n_2/p$; see Fig.~\ref{fig:sliced_pencils}. The value of $p$ is chosen so as to capture this inhomogeneity, while retaining a balance between computation and communication. This \emph{sliced pencils} approach for the distributed FFT-accelerated Poisson equation is summarized in Algorithm~\ref{alg:fast_poi_solver_par}.
  \item[--] \emph{\textbf{The slab-decomposed approach.}} Finally, we devised an alternative approach to solve the decoupled 2D systems at the cost of one \emph{all-to-all} collective operation. The approach follows the computation of the FFT-based transforms by a pencil--slab data redistribution, allowing to solve the $2D$ systems explicitly, with balanced loads. For the sake of conciseness, this approach is described in \ref{sec:poi_other}.
\end{itemize}
\begin{algorithm}[hbt!]
  \caption{Sequence of operations performed per task $m$ for the parallel solution of the Poisson equation (Eq.~\eqref{eqn:poi}) with FFT-based synthesis, using ``sliced pencils''; see Fig.~\ref{fig:sliced_pencils}.}\label{alg:fast_poi_solver_par}
  \begin{algorithmic}
    \setstretch{1.0}
    \small
    \State \textbf{do} {$i=1$ to $n_1^m$ and $k=1$ to $n_3^m$}
    \State \quad forward FFT-based transform along $x_2$ of right-hand-side of Eq.~\eqref{eqn:poi}: $\hat{f}_{i,1 \dots n_2,k} = \mathcal{F}_{x_2}(f_{i,1 \dots n_2,k})$
    \State \textbf{end do}
    \State \textbf{do} $J=1$ to $p$ \Comment{$p$, 3D problems decoupled along $x_2$}
    \State \quad solve Eq.~\eqref{eqn:poi_reduced} in pencil chunk \verb|J| within \verb|comm_world| using a geometric multigrid solver, to obtain $\hat{\Phi}_{1 \dots n_1^m,(J-1)n_2^p+1 \dots J n_2^p,1 \dots n_3^m}$
    \State \textbf{end do}
    \State \textbf{do} $i=1$ to $n_1^m$ and $k=1$ to $n_3^m$
    \State \quad backward FFT-based transform along $x_2$ of the solution: $\Phi_{i,1 \dots n_2,k} = \mathcal{F}_{x_2}^{-1}(\hat{\Phi}_{i,1 \dots n_2,k})$
    \State \textbf{end do}
  \end{algorithmic}
\end{algorithm}\par
\section{Validation and Computational performance}\label{sec:results}
\subsection{Validation}\label{sec:validation}
Before presenting the validations of the numerical algorithm, we should note that verifying the implementation of the Poisson solver and pressure projection steps is simple, as the final velocity has to be divergence-free (up to the tolerance conditioned by the iterative error). This incompressibility condition is checked recurrently during the calculation.\par
Besides the different solution approach for the Poisson equation, the numerical method is equivalent to that of \emph{CaNS}, which has been validated against several canonical turbulent flows (e.g.\ channel, square duct, and decaying Taylor-Green vortex) \cite{Costa-CAMWA-2018}. Hence, for simple rectangular boxes, all the validations shown in \cite{Costa-CAMWA-2018} for turbulent flows are easily reproduced by the present tool. We therefore restrict ourselves to computationally cheaper test cases in multi-block geometries. Unless otherwise stated, the simulations are integrated in time with a varying time step, $dt = \mathrm{CFL}\,dt_{max}$, with $dt_{max}$ the maximum allowed time step, and $\mathrm{CFL}=0.95$; the PFMG solver was seen to be efficient and robust enough for all cases, with tolerance and maximum number of iterations set to $10^{-4}$ and $50$. Hereafter, $u$, $v$, and $w$ will denote the $x$, $y$, and $z$ components ($x_1$, $x_2$, and $x_3$ above) of the velocity. Finally, we should note that, for the same ``assembled'' computational setup, the numerical results should be independent of the block and MPI partitioning, to machine precision.
\subsubsection*{Three-dimensional lid-driven cavity flow}
We consider a three-dimensional lid-driven cavity flow, simulated in a cubic domain with dimensions $[-H/2,H/2]^3$. Zero velocity boundary conditions are prescribed at all the boundaries, except for the top wall, which moves with a velocity $u(x, H/2, z)=(U_L,0,0)$; the Reynolds number is $\mathrm{Re}=U_LH/\nu=1000$, and the flow is solved on a uniform grid with spacing $\Delta \ell = H/128$. \par
\begin{figure}[hbt!]
  \centering
  \includegraphics[width=0.45\columnwidth]{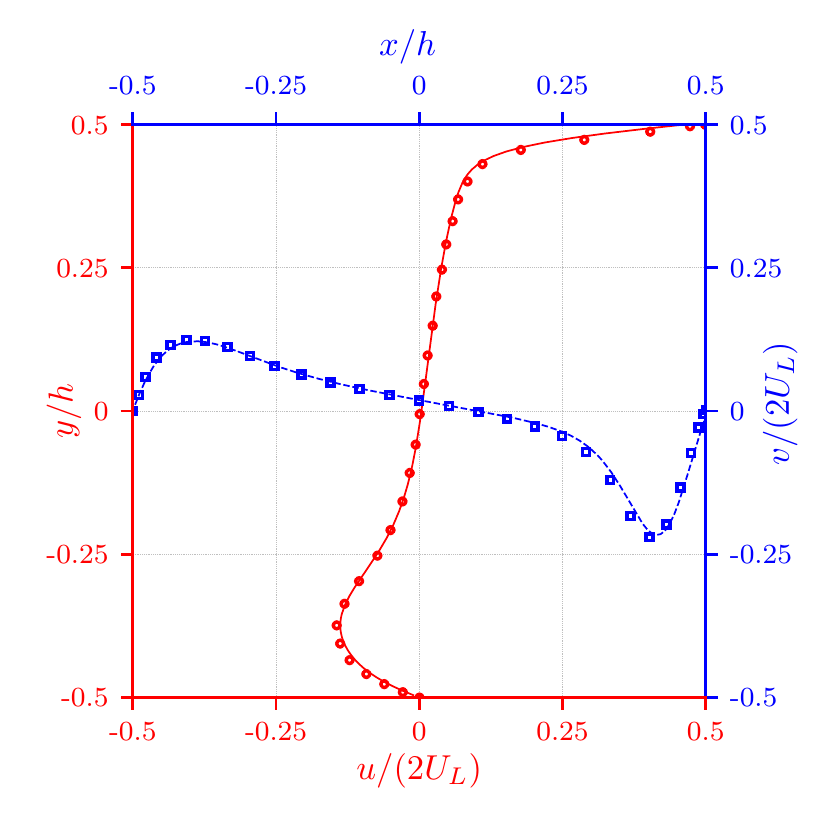}
  \caption{Normal velocity profiles along the centerlines $u(0,y,0)$ and $v(x,0,0)$ for a lid-driven cubic cavity at $\mathrm{Re} = 1000$. The symbols correspond to the data extracted from \cite{Ku-et-al-JCP-1987}.}\label{fig:valid-ldc}
\end{figure}

Fig.~\ref{fig:valid-ldc} shows the velocity profiles of the steady-state solution at the centerlines $u(0,y,0)$ and $v(x,0,0)$, compared to the data extracted from \cite{Ku-et-al-JCP-1987}, showing good agreement. It should be noted that the same setup was validated in \cite{Costa-CAMWA-2018}, and the present results match that data with a maximum relative difference of $10^{-7}$. We have also confirmed that partitioning the geometry into smaller individual blocks (e.g.\ six, two per domain direction) results in the exact same calculation.
\subsubsection*{Laminar flow through a T-junction}
\begin{figure}[hbt!]
  \centering
  \includegraphics[width=0.80\columnwidth]{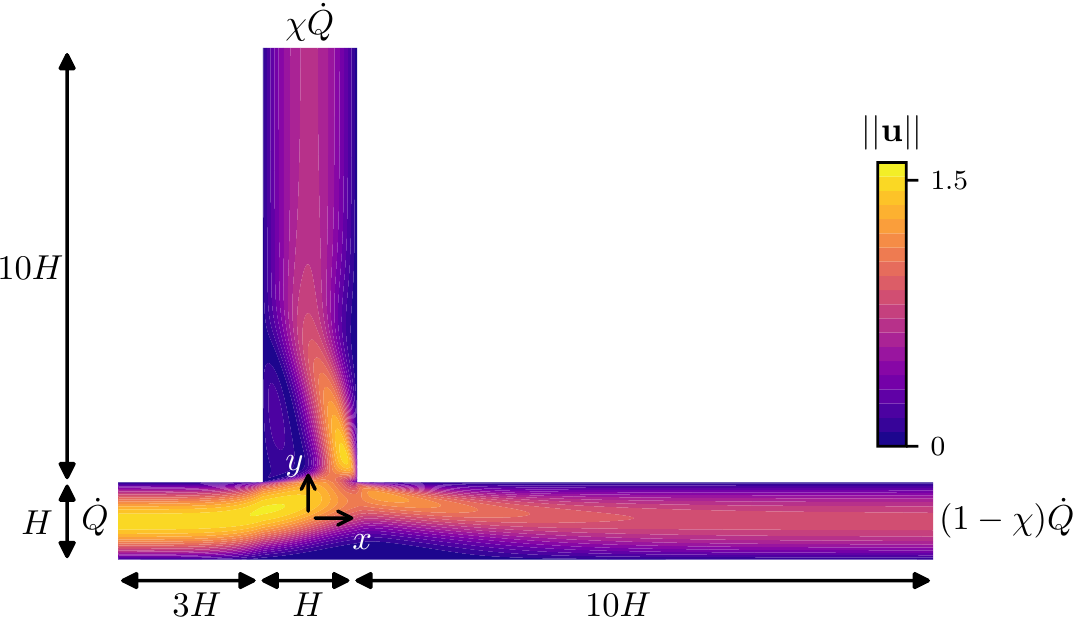}
  \caption{Schematic of the configuration for the T-junction simulation, with the contours of the steady-state velocity magnitude. A fully-developed Poiseuille profiles are prescribed with a flow rate $\dot{Q}$ (inlet), $(1-\chi)\dot{Q}$ (main branch outlet), and  $\chi\dot{Q}$ (derivative branch outlet). Note that the height of the channel was increased for clarity.}\label{fig:valid-tj}
\end{figure}

We simulated the laminar T-junction flow shown in Fig.~\ref{fig:valid-tj}, with a constant channel height $H$, and composed of a short inlet branch, and two longer outlet branches, a geometry which requires at least four distinct blocks (cf.\ Fig.~\ref{fig:illustration_block}). A fully developed Poiseuille profile is prescribed at the inlet, corresponding to a flow rate per unit depth $\dot{Q}$. At the outlet, the same profiles are prescribed, but for an exiting flow rate of $\chi\dot{Q}$ in the branching (vertical) channel, and $(1-\chi)\dot{Q}$ in the main (horizontal) channel, with $\chi=0.44$; no-slip and no-penetration boundary conditions are prescribed at the walls. The flow is governed by a Reynolds number $\mathrm{Re} = \dot{Q}/\nu=248$, and is solved on a regular grid with constant spacing, $\Delta \ell = H/64$.
The steady-state solution is depicted in Fig.~\ref{fig:valid-tj}, showing the velocity magnitude.
\par
\begin{figure}[hbt!]
  \centering
  \includegraphics[width=0.42\columnwidth]{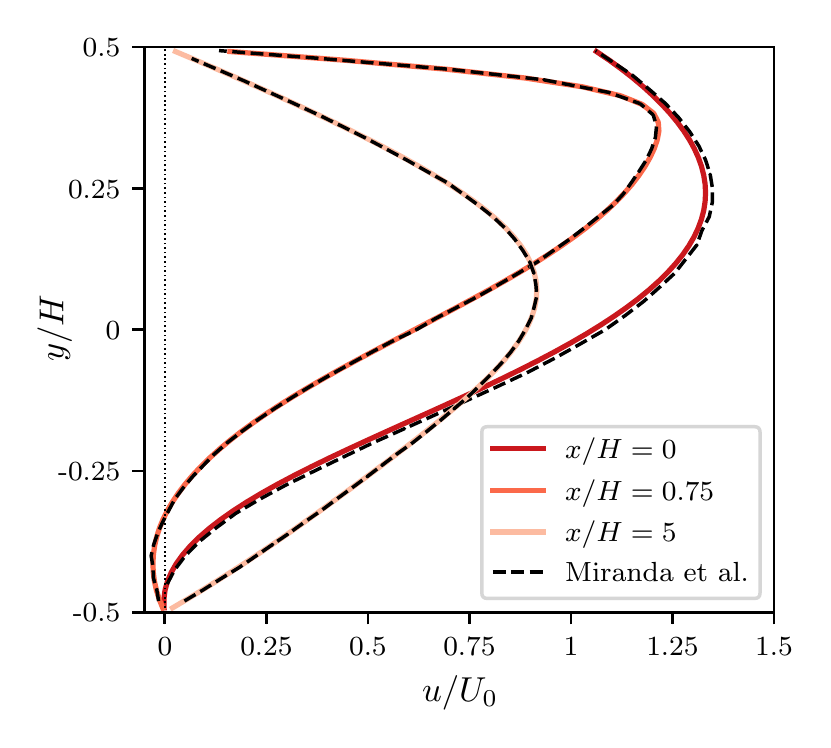}\hfill
  \includegraphics[width=0.42\columnwidth]{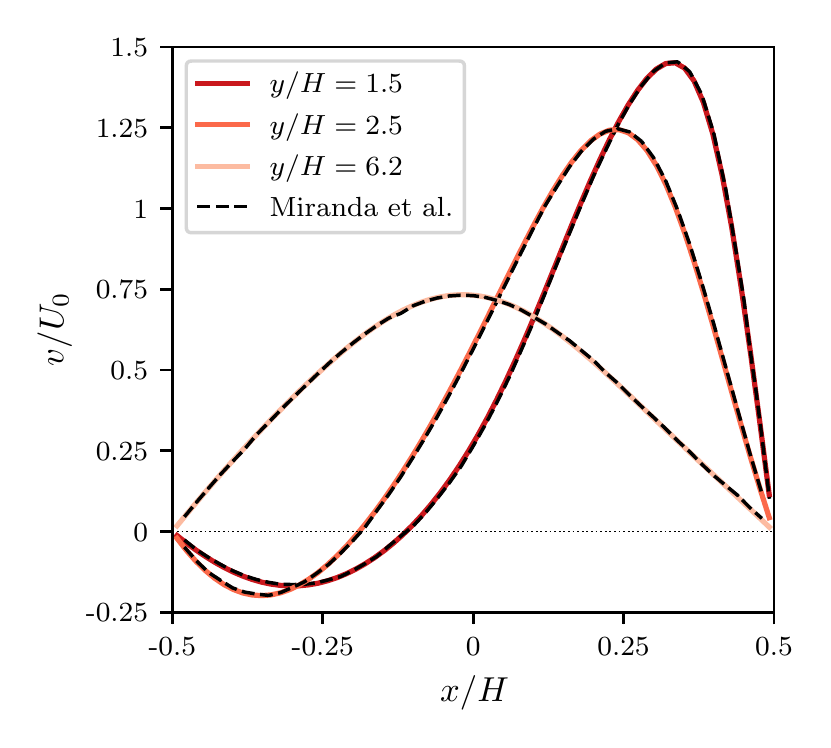}
  \caption{Profiles of streamwise velocity in the main branch (left) ($x$ component, $u$), and derivative branch (right) ($y$ component, $v$) of a T-junction, compared to the reference data in \cite{Miranda-et-al-IJNMF-2008}. $U_0=\dot{Q}/H$ is the inlet bulk velocity.}\label{fig:valid-tj-profiles}
\end{figure}

This computational setup was studied numerically for Newtonian and non-Newtonian fluids in Ref.~\cite{Miranda-et-al-IJNMF-2008}, to reproduce the experiments in Ref.~\cite{Liepsch-et-al-JoB-1982}. The shape and extent of the two recirculation regions at the entrance of each branch agree with what is reported in these references. More quantitatively, Fig.~\ref{fig:valid-tj-profiles} shows the profiles of streamwise velocity in the main branch and derivative branches, at different cross-sections, compared to the reference data extracted from \cite{Miranda-et-al-IJNMF-2008}. The agreement is excellent.
\subsection{Computational performance}\label{sec:performance}
We now assess the performance of the numerical algorithm in massively parallel calculations, with the different approaches for solving the Poisson equation. For convenience, we will use the following notation for the different approaches:
{
\begin{itemize}
  \itemsep0em
  \item[--] \textbf{\nofft:} iterative solution of a single 3D system without FFT acceleration (\S\ref{sec:plain_3d_mg});
  \item[--] \textbf{\yofft:} iterative solution of 2D systems in a pencil decomposition, after FFT synthesis (i.e., the maximum partition of case \yrfft, but using 2D matrices) -- the naive approach in \S\ref{sec:mg_w_fft};
  \item[--] \textbf{\yrfft:} iterative solution of $p$, 3D systems over partitioned pencils, decoupled along the FFT direction, after FFT synthesis (Algorithm~\ref{alg:fast_poi_solver_par}) -- the sliced pencils approach in \S\ref{sec:mg_w_fft};
  \item[--] \textbf{\yefft:} iterative solution of 2D systems in a slab decomposition, after FFT synthesis and a pencil--slab data redistribution (Algorithm~\ref{alg:fast_poi_solver_par_oth}) -- the slab-decomposed approach in \S\ref{sec:mg_w_fft}.
\end{itemize}}\par
\begin{figure}[hbt!]
  \centering
  \includegraphics[height=0.33\columnwidth]{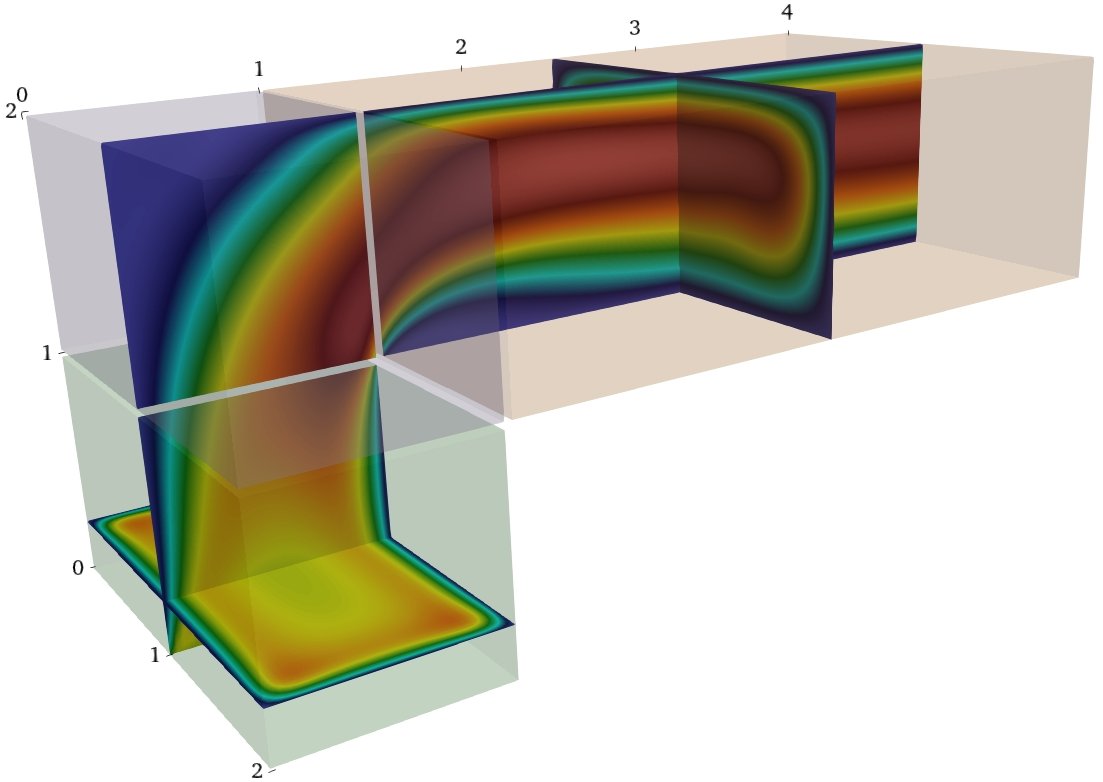}\hfill
  \includegraphics[height=0.52\columnwidth]{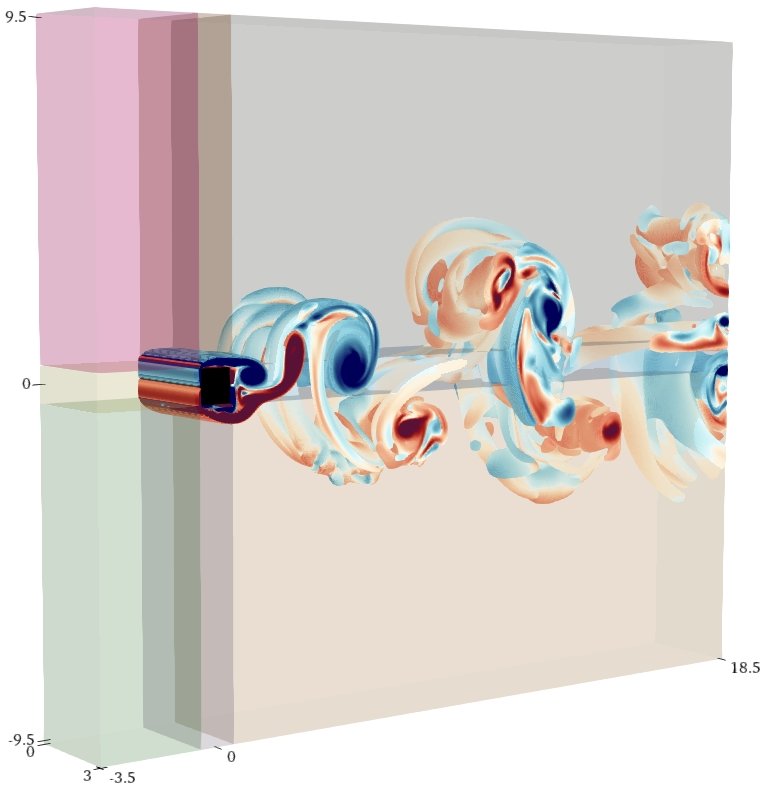}
  \put(-100,15 ){\scriptsize$x/H$}
  \put(-242,122){\scriptsize$y/H$}
  \put(-225,5  ){\scriptsize$z/H$}
  \put(-380,152){\scriptsize$x/H$}
  \put(-475,80 ){\scriptsize$y/H$}
  \put(-453,18 ){\scriptsize$z/H$}
  \put(-280,5){\small(a)}
  \put( -30,5){\small(b)}
  \caption{Multi-block computational setups considered in the scaling performance assessment, with the domain dimensions and the blocks depicted in different translucent colors. (a): flow in a L-shaped rectangular duct with height $H$, solved on a constant grid with $H/\Delta \ell=256$. No-slip boundary conditions are prescribed everywhere, except at the inflow (bottom, with uniform velocity $U$), and at the outflow (zero pressure); the Reynolds number is $\mathrm{Re}=UH/\nu=500$. The planar contours show the steady-state velocity magnitude (red -- high; blue -- low). (b): turbulent flow around a square cylinder with size $H$ solved on a constant grid with $H/\Delta \ell=64$. The flow is periodic along the $z$ direction, and a uniform velocity $U$ is prescribed at the inflow, with zero pressure boundary conditions prescribed elsewhere for simplicity; the Reynolds number is $\mathrm{Re}=500$. The figure shows the regions of the domain with vorticity $|\boldsymbol{\omega}|>5U_b/H$, colored by the local spanwise vorticity $\omega_z$ in a divergent linear colormap (blue to red) clamped at $\omega_z=\pm3$. For the performance tests, the size of the top, bottom, and right blocks was divided by $3$, to allow assessing scaling over a range of $O(10)-O(1000)$ cores.}\label{fig:visu_3d_cyl}
\end{figure}
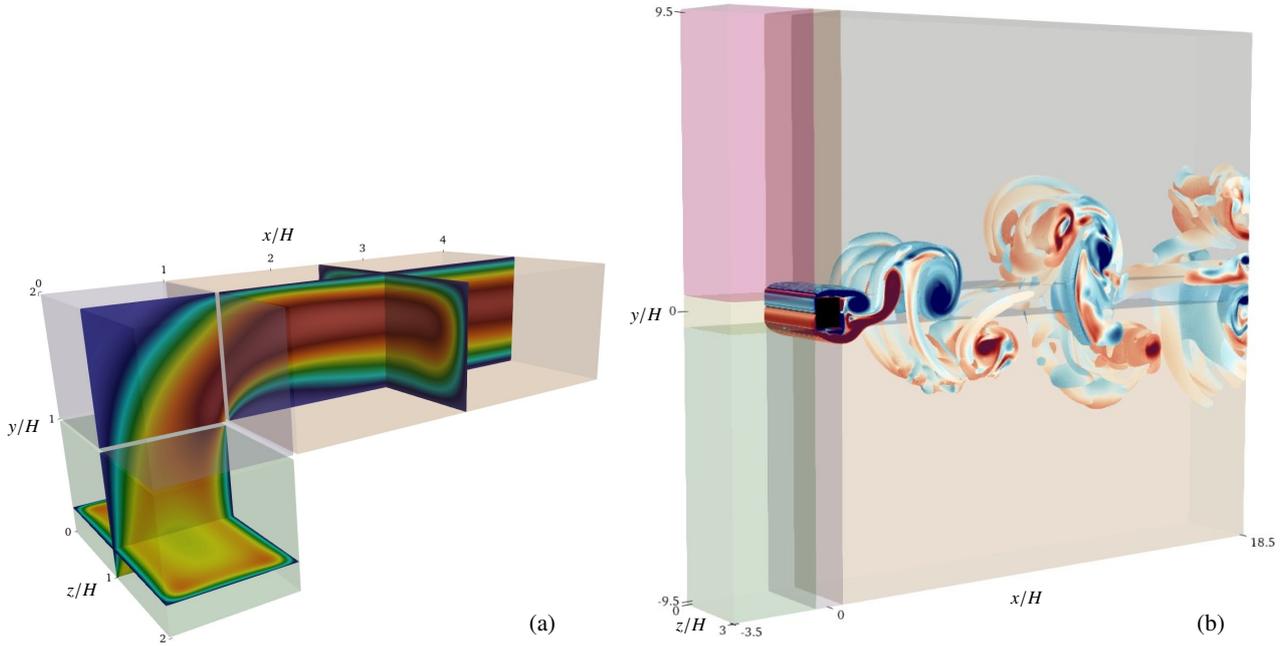

Three different setups are considered, with geometries defined by an increasing number of blocks: a lid-driven cavity flow ($1$ block), an L-shaped duct ($3$ blocks), the flow around a square obstacle ($8$ blocks). The lid-driven cavity flow corresponds to the problem described in \S\ref{sec:validation}; the other two cases are illustrated in Fig.~\ref{fig:visu_3d_cyl}, where the block partitioning can be also appreciated, and the computational parameters are described in the figure caption. Note that the L-channel is an example of a system possibly better suited for a multi-block solver than a single-block DNS solver extended with an immersed boundary method. Conversely, the flow around a square setup is more suited for leveraging such a single-block approach (see \cite{Chiarini-and-Quadrio-FTC-2021}), because it can be represented by a rectangular box with only a small portion of the domain -- the square obstacle -- excluded.\par
The timing measurements reported here correspond to the wall-clock time required to perform a full solution time step (i.e., three RK3 substeps), averaged over $100$ instances. As we will see, the majority of this time is spent solving the Poisson equation, roughly $85\%-95\%$, depending on the approach. The runs were performed on the supercomputer Tetralith based in Sweden (Xeon Gold 6130 16C 2.1GHz, Intel Omni-Path), with \emph{SNaC} built using the Intel programming environment (18.0.1) with \verb|-O3 -fp-model fast -xHost| as optimization flags. For all the cases here, a pencil partitioning $p=16$ (recall Algorithm~\ref{alg:fast_poi_solver_par}) will be used, as it was found to result in a good scaling performance. A more detailed analysis of the influence of this parameter in the algorithm performance is presented in \ref{sec:pencil_slicing_analysis}.\par
\par

\begin{figure}[hbt!]
  \centering
  \includegraphics[width=0.49\columnwidth]{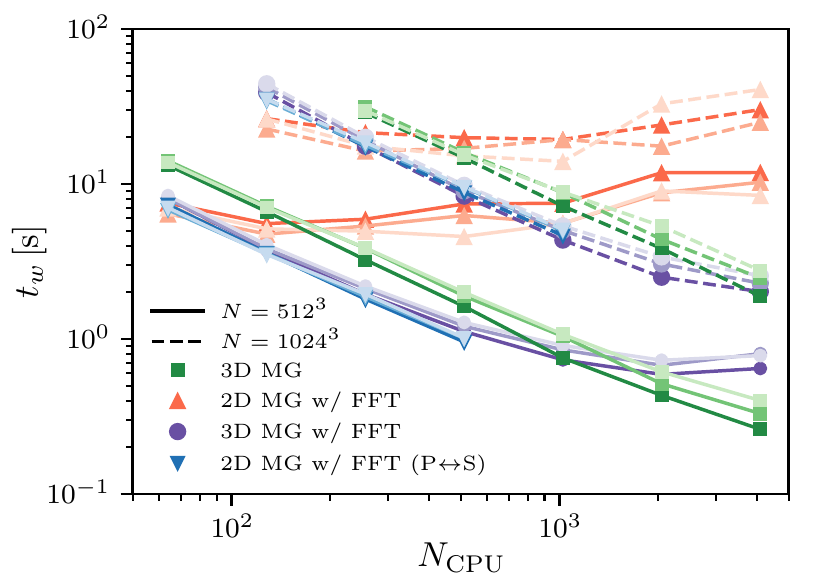}\hfill
  \includegraphics[width=0.49\columnwidth]{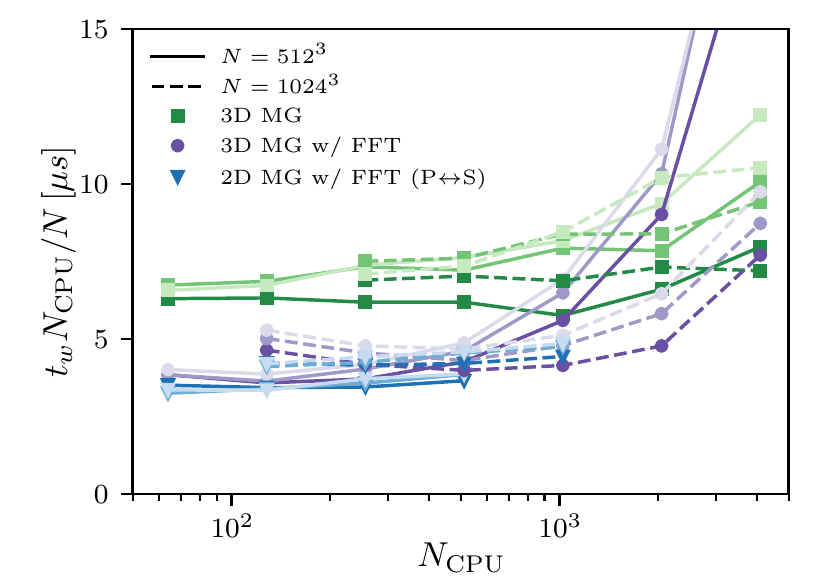}
  \put(-260,5){\small(a)}
  \put( -20,5){\small(b)}
  \caption{(a): strong scaling of the numerical algorithm up to $4096$ cores, for a lid-driven cavity flow with $512^3$ and $1024^3$ grid cells, with different directions of FFT-based synthesis ($x$-, $y$-, or $z$-oriented subdomains by increasing color lightness). $t_{w}$ denotes wall-clock time in seconds/time step/task (i.e., three Runge-Kutta substeps), and $N_{CPU}$ the number of cores. (b): compensated scaling plot showing the total CPU time per grid cell, per time step, with $N$ being the total number of grid cells. Here a horizontal line corresponds to ideal scaling. Legend notation -- see the beginning of \S\ref{sec:performance}.}\label{fig:scaling_ldc}
\end{figure}

Fig.~\ref{fig:scaling_ldc}(a) shows the strong scaling performance of the single-block case (lid-driven cavity) for two different grids (with $N=512^3$ and $1024^3$), with different directions of FFT synthesis. The differences in performance for the different pencil orientations are small, with $x$-aligned pencils performing slightly better for the \nofft case, possibly due to a more favorable data distribution; note that, since the grid is constant, the PFMG solver should coarsen along $x$ in this problem. Interestingly, when FFT synthesis is used, the timings are much less sensitive to the pencil orientation. As expected from the excellent performance of the \emph{hypre} library, the geometric multigrid solver without FFT acceleration scales very well, as it can be also depicted in the compensated plot in panel (b) of Fig.~\ref{fig:scaling_ldc}. Note that, there, the slight offset between cases with $512^3$ and $1024^3$ is due to a slightly larger number of iterations required for the iterative solver on the finer grid.\par
Somewhat expectedly, the FFT-accelerated approaches perform well for a small number of cores, showing a remarkable $2$-fold speedup compared to the standard 3D multigrid solution. However, when increasing the number of cores, the importance of solving several 2D systems in parallel becomes evident. While the scaling quickly degrades when the 2D linear systems are solved naively in the pencil decomposition (\yofft), it remains excellent with the other two approaches: when the slab-decomposed solution is used (\yefft; Algorithm~\ref{alg:fast_poi_solver_par_oth}) the figure shows a consistent $2$-fold speedup, until the maximum partitioning is reached; using the sliced pencils approach (Algorithm~\ref{alg:fast_poi_solver_par}; \yrfft) shows similar performance, but allows to reach a higher number of cores, until the load per task becomes too small and the scaling deteriorates. This occurs for a number of cores $N_{CPU}$ beyond $1024$ for the $512^3$ setup, and beyond $2048$ for the $1024^3$ case. Nevertheless, the wall-clock time per step in the scaling region is already quite small.\par
These figures are expected to scale to larger, more ambitious, problem sizes. To illustrate this, we also investigated the weak scaling performance of the same problem, and performed a strong scaling analysis on a different machine at more extreme scales -- up to $65\,536$ cores for a domain with $4096^3$ grid points. For the sake of conciseness, these results are discussed in \ref{sec:scaling_big}.\par
\par
To breakdown the different contributions of the calculation timeline to the total wall-clock time, we profiled the application\footnote{{Using the Arm~MAP profiler~21.0.2.}} for two computational grids assessed in Fig.~\ref{fig:scaling_ldc}: a $512^3$ box decomposed among $64$ CPUs, and a $1024^3$ box decomposed among $1024$ CPUs; note that $N/N_{CPU}$ varies by a factor of two between the cases. Unfortunately, the profiling overhead resulted in a performance degradation which was disproportionately larger for cases \yrfft and \yofft, especially for many-core runs with a small number of points per task. Hence, we restricted the analysis to setups with a substantial amount of grid points per task. Moreover, the naive approach (\yofft) metrics were severely exacerbated by the profiling at all scales analyzed, and the results are therefore not shown.\par
In addition to the performance metrics we present below, the profiler also measured the memory footprint of the different approaches (note that double precision is used). The cases \nofft and \yrfft used roughly $480$ bytes per grid point, slightly more than the other two approaches which solve explicitly 2D systems, \yefft and \yofft, which used about $350$ bytes per grid point.\par
\begin{figure}[hbt!]
  \centering
  \includegraphics[width=0.99\columnwidth]{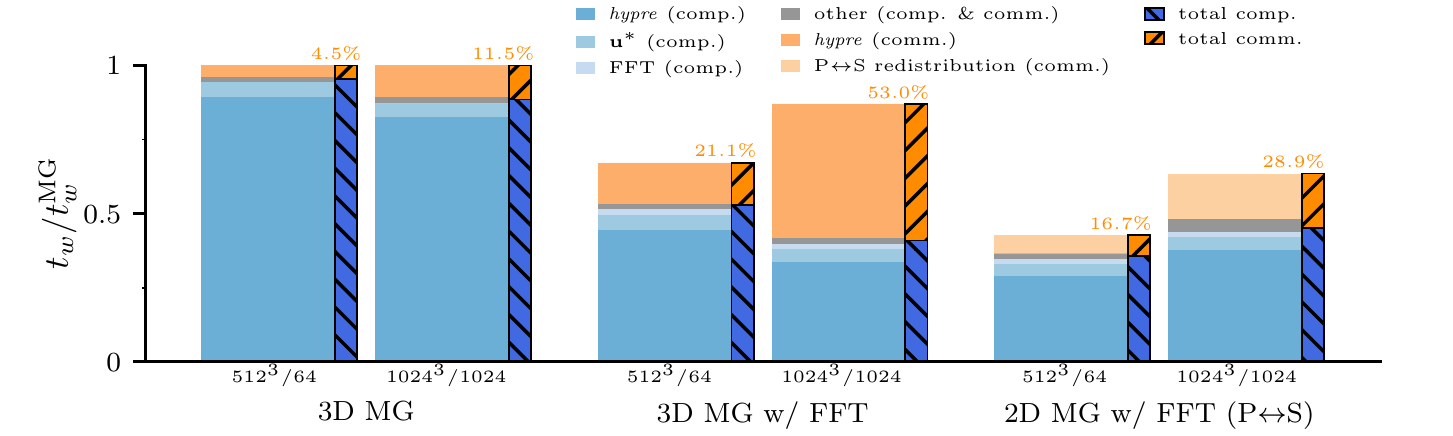}\hfill
  \caption{Breakdown of the different shares of the calculation timeline to the total compute time per time step, for a single-block calculation (lid-driven cavity case) and increasing domain decomposition. The bars show the wall-clock time per step, $t_w$, normalized by that of the \nofft case, $t_w^\mathrm{MG}$, for the three different approaches denoted in the horizontal axis, and two domain decompositions: $N/N_{CPU} = 512^3/64$ and $1024^3/1024$. Blue shaded areas denote computation operations as described in the legend, while orange ones denote communication. In the legend, $\mathbf{u}^*$ denotes all operations required for the calculation of the prediction velocity. The gray area denotes the remainder of the share of calculation time (communication and computation). The numbers on top of each bar correspond to the communication share of the total calculation time.}\label{fig:scaling_profiling}
\end{figure}

These results are plotted in Fig.~\ref{fig:scaling_profiling}, where the bars show the calculation wall-clock time $t_w$, normalized by that of the \nofft case, $t_w^\mathrm{MG}$. Expectedly, the relative communication footprint increases with increasing decomposition, and is larger for the cases which exploit FFT acceleration, as they are computationally cheaper. Also not surprisingly, the computation footprint of calculating the prediction velocity $\mathbf{u}^*$ is the same among cases, and ditto for the FFT-based transforms, for the cases which exploit them. Interestingly, virtually all the communication is associated with the solution of the Poisson equation, meaning that the overhead associated with the halo exchanges is quite small. The solution of the Poisson equation takes no less than roughly $90\%$ of the calculation time, but is much smaller for the cases with FFT synthesis. Of course, the breakdown of the different contributions to the Poisson solver footprint is also quite different. As expected, performing the FFT-based acceleration of the Poisson equation results in a significant speedup of up to a factor two for this case, with a quite small overhead to compute the FFT-based transforms (here, cosine transforms), taking no more than $2\%$ of the total time. Note also that the communication operations performed within \emph{hypre} are suppressed for case \yefft, and replaced by those of the pencil--slab data redistribution. This is expected: for a single-block calculation, the pencil--slab data redistribution will serialize the 2D multigrid solves, because each slab contains a batch of undivided 2D problems. Conversely, in a multi-block setup, the 2D multigrid solves following the pencil--slab data redistribution will still require inter-block communication, as Fig~\ref{fig:transpose_to_slab} illustrates. We should finally note that, when profiling, the timings for case \yefft appear to be much smaller than that of \yrfft, contrary to what Fig.~\ref{fig:scaling_ldc} shows. The reason is that the timings for case \yrfft were more penalized by the profiler.\par
\par
\begin{figure}[hbt!]
  \centering
  \includegraphics[width=0.49\columnwidth]{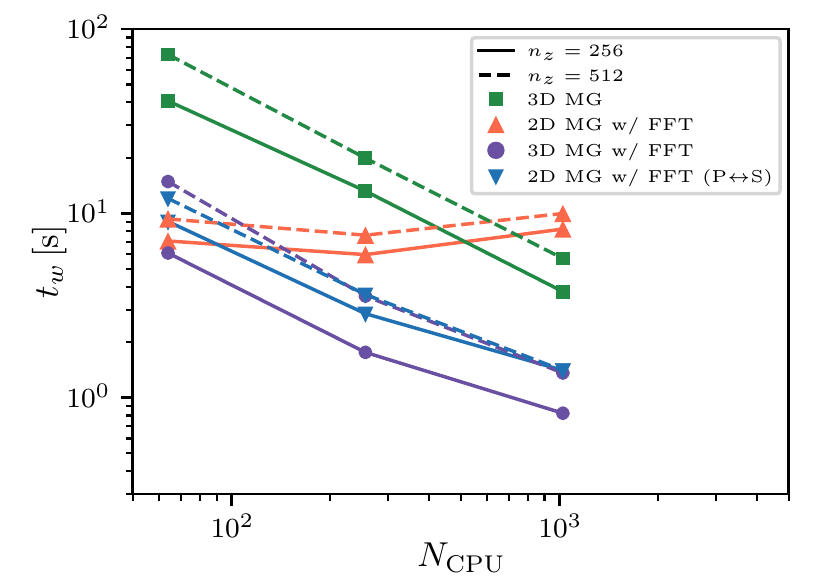}\hfill
  \includegraphics[width=0.49\columnwidth]{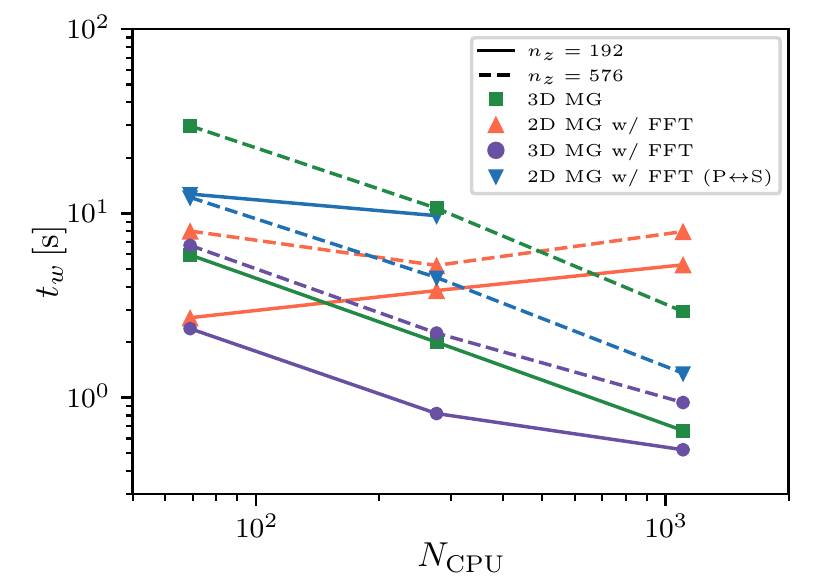}
  \put(-260,5){\small(a)}
  \put( -20,5){\small(b)}
  \caption{Strong scaling of the numerical algorithm for the L-channel case (a), and flow around a square case (b), plotting the same quantities as Fig.~\ref{fig:scaling_ldc}(a). The different approaches are considered, using $z$ as the direction of FFT synthesis, for two different values of the number of grid points $n_z$. Legend notation -- see the beginning of \S\ref{sec:performance}.}\label{fig:scaling_other}
\end{figure}

Fig.~\ref{fig:scaling_other} shows the strong scaling performance of the other two cases considered, with a $z$-aligned pencil decomposition, and two different values of $n_z$ while keeping the number of points in the other directions fixed; the domain length along $z$ was also increased to keep the grid spacing constant. The blocks were decomposed among MPI tasks with a constant number of grid points per computational subdomain, to ensure load balancing. The only exception is the right-most block in the L-channel (Fig.~\ref{fig:visu_3d_cyl}), which was less decomposed (a factor $1.5$ more grid points per task than the other blocks), so that we could still test Algorithm~\ref{alg:fast_poi_solver_par_oth} in $O(1000)$ tasks without adding more spanwise grid points. Moreover, this allows assessing the performance of a setup with a small load imbalance.\par
Remarkably, FFT-based acceleration results in a tremendous speedup for the L-channel case, with an almost $8$-fold speedup compared to the standard iterative solution. Here, for the smaller value of $n_z=256$, the sliced pencils approach in Algorithm~\ref{alg:fast_poi_solver_par} performs best. Conversely, with larger values of $n_z$ the overhead of the \emph{all-to-all} collective in the slab-decomposed approach (Algorithm~\ref{alg:fast_poi_solver_par_oth}) becomes less significant, and the two approaches show very similar performance. Despite these differences, both approaches show a remarkable speedup, allowing for very small values of wall-clock time per step. We should note that, despite the large speedup in the L-channel case, the best wall-clock time is still larger (by roughly a factor of $2$) than the single-box solver \emph{CaNS} in a box that fits the L-channel. As a rough estimate, we expect savings in wall-clock time when the multi-block calculation requires about $3.5$ fewer grid cells than the corresponding single-box envelope. Of course, the fast single-box solver cannot exploit non-uniform grids along more than one direction, and the imposition of boundary conditions at immersed boundaries is not exact.\par
Conversely, for the flow around the square case, the performance of the FFT-accelerated solver is less impressive, because the value of $n_z$ relative to the problem size is smaller. Nevertheless, for larger $n_z$, up to about $5$-fold speedup can be observed. Here the communication overhead of the slab decomposed solver is too large, resulting in relatively poor performance. Still, despite the reasonable performance here for a smaller number of cores, we recall that this case may be more suited for a simpler, single-box solver extended with an immersed boundary method.\par
\begin{figure}[hbt!]
  \centering
  \includegraphics[width=0.99\columnwidth]{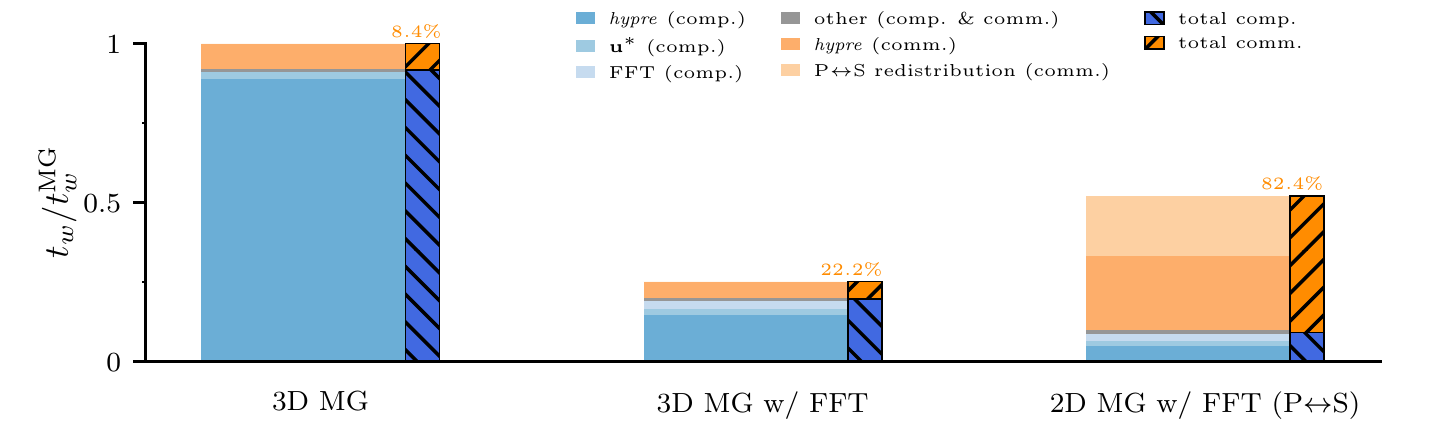}\hfill
  \caption{Breakdown of the different shares of the calculation timeline to the total compute time per time step, for a multi-block calculation (flow around the square case, with $n_z=576$), on $N_{CPU}=69$ cores. The communication footprint in the \yefft case has been slightly penalized by the profiling. See the legend of Fig.~\ref{fig:scaling_profiling} for more details.}\label{fig:scaling_profiling_fas}
\end{figure}

To highlight the performance of the different approaches in a multi-block setting, Fig.~\ref{fig:scaling_profiling_fas} presents the profiling results for the flow around the square case\footnote{Unfortunately, we were unable to obtain reliable the profiling results for the L-channel case, since the metrics were substantially penalized by the profiler, possibly due to the prescribed load imbalance in this setup.}. Clearly, the sliced pencils approach is superior in terms of wall-clock time and communication overhead. Of course, due to the large speedup, its relative communication footprint seizes a larger share of the total calculation time compared to case \nofft, about $20\%$. Finally, the slab-decomposed (\yefft) case shows a huge communication overhead. While these metrics are possibly aggravated by the profiling itself, the difference in performance compared to the single-block profiling in Fig.~\ref{fig:scaling_profiling} are expected: in a single block, the pencil--slab data redistribution serializes the subsequent iterative solutions, meaning that there should be no communication within the \emph{hypre} library; in a multi-block setting, inter-block communications are required (see Fig.~\ref{fig:transpose_to_slab}), and the footprint of the communication within \emph{hypre} is also significant. Finally, it is worth noting that the computation footprint is the smallest for the \yefft case. This is expected, because this approach perfectly covers the inhomogeneity of the iterative systems along the synthesis direction (see the analysis in \ref{sec:pencil_slicing_analysis}). However, compared to case \yrfft, the communication cost of this approach clearly outweighs the benefit of the optimal coverage of the problem inhomogeneity.\par
\begin{figure}[hbt!]
  \centering
  \includegraphics[width=0.85\columnwidth]{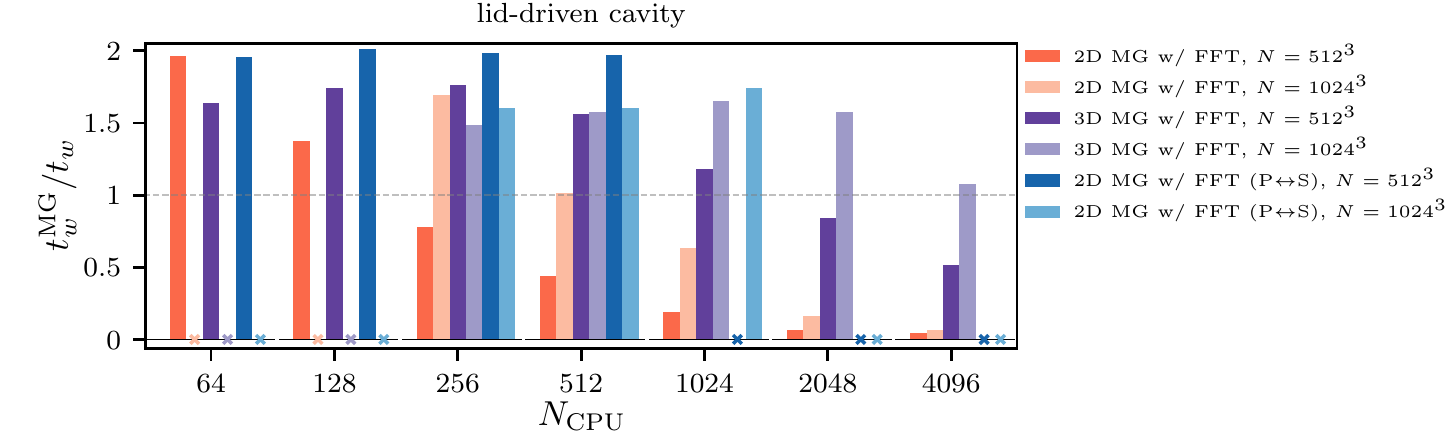}\vspace{0.5cm}
  \includegraphics[width=0.85\columnwidth]{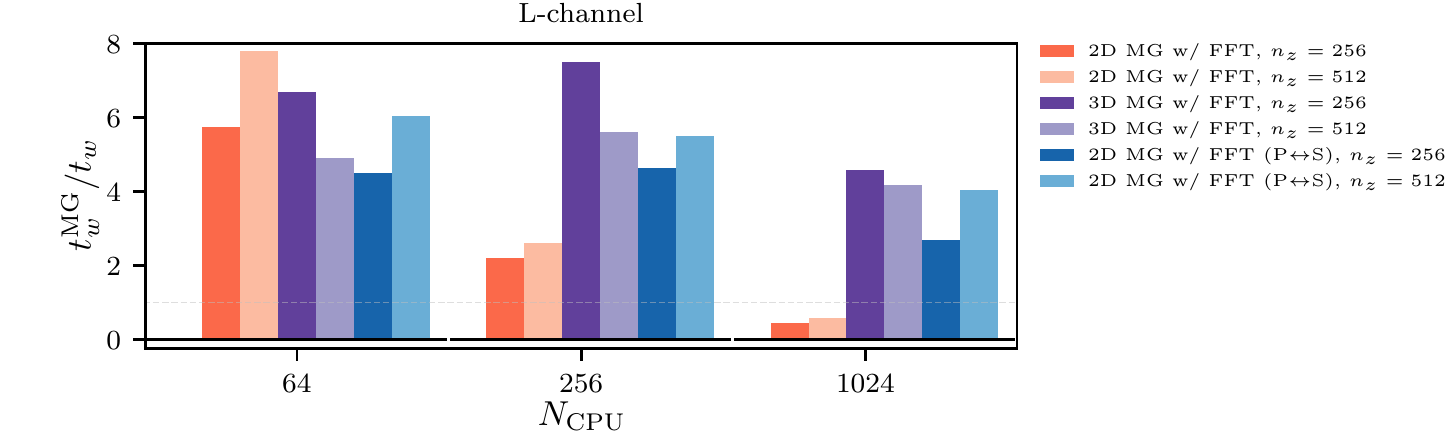}\vspace{0.5cm}
  \includegraphics[width=0.85\columnwidth]{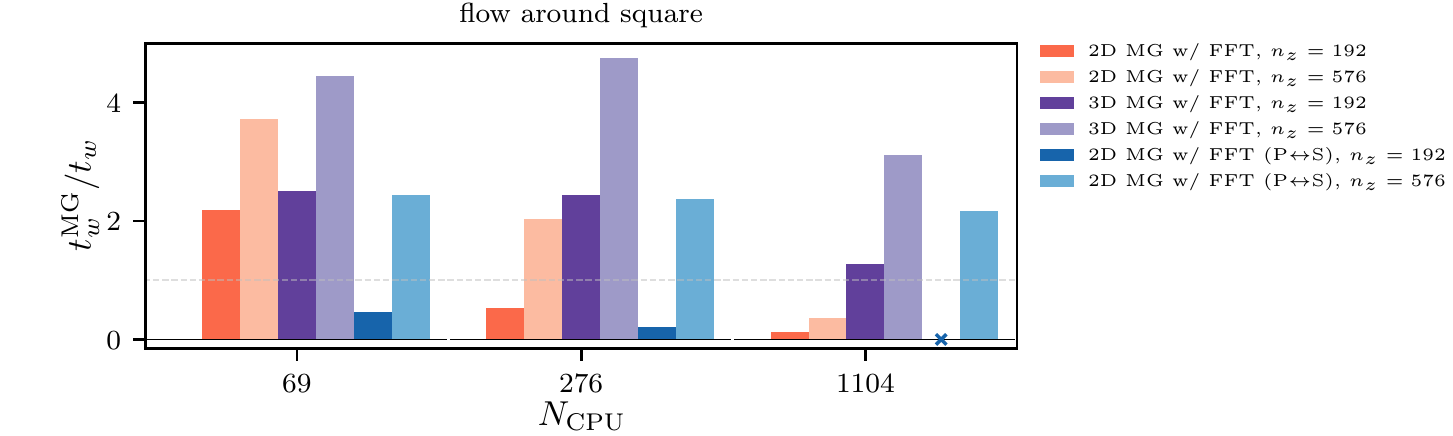}
  \caption{Speedup of the FFT-accelerated calculation with respect to the standard multigrid solution (with wall-clock time per step $t_w^\mathrm{MG}$), as a function of the number of cores $N_{CPU}$, for the different cases considered, with $z$ is taken as the FFT synthesis direction. The grey dashed line marks the threshold of performance gain $t_w=t_w^\mathrm{MG}$, and the ``$\boldsymbol{\times}$'' markers denote simulations that were not possible to perform due to insufficient memory (for small number of cores), or not enough points to slab-decompose along $z$ (for larger number of cores in case \yefft). Legend notation -- see the beginning of \S\ref{sec:performance}.}\label{fig:speedup_all}\par
\end{figure}

To get a better impression of the performance gains for these three different canonical systems, Fig.~\ref{fig:speedup_all} summarizes the increase in wall-clock time per step of the FFT-accelerated calculation, relative to the standard iterative solution. Clearly, the method performs best when the number of points in the direction of FFT synthesis is larger, which ensures a substantial load per task. Nonetheless, the results demonstrate the potential of this approach to speedup a multi-block DNS by large factors, and with small enough wall-clock time per time step. On balance, the best-performing approach is clearly the ``sliced pencils'' one (Algorithm~\ref{alg:fast_poi_solver_par}; \yrfft).
\section{Summary and Outlook}\label{sec:conclusions}
We have presented and validated a fast and versatile multi-block finite-difference solver for the incompressible Navier-Stokes equations. If the physical problem features one homogeneous direction, which is the case in numerous setups of interest, the numerical algorithm can exploit the method of eigenfunctions to decouple the finite-difference Poisson equation along that direction. This ``synthesis'' of the Poisson equation can be employed at a very low cost using FFT-based transforms, and enables major gains in the performance of the overall numerical algorithm. We have implemented the different FFT-based expansions in a unified framework, to support all the valid combinations of boundary conditions of the method.\par
Following the FFT-based synthesis, a series of two-dimensional Poisson problems are solved using an efficient geometric multigrid solver. Here we leveraged the well-established \emph{hypre} library, which enables a flexible multi-block implementation, however with excellent performance. We have demonstrated that the most straightforward application of the library to this problem is bound to show poor parallel performance, and proposed two distinct strategies to improve the parallel scalability of the overall method. Both strategies were shown to greatly improve the parallel performance of the algorithm, allowing for $2$- to $8$-fold speedups of the calculation, corresponding to a small wall-clock time per time step. However, one of these stood out, by exploring an optimal trade-off between capturing the inhomogeneity of the 2D problems in the FFT direction, and maintaining a significant compute load per task. This approach was shown to perform well for all configurations, and in a truly massively parallel setting, scaling at least up to $65\,536$ cores.\par
The numerical algorithm was implemented in a new DNS code, \emph{SNaC}, which was made freely available and open-source. Given the flexibility and great performance of the tool, \emph{SNaC} is expected to follow the footsteps of other research DNS codes such as \emph{CaNS} and \emph{AFiD}, and serve well as a base multi-block Navier-Stokes solver on top of which approaches for more complex phenomena can be implemented, such as immersed boundary methods for complex geometries \cite{DallaBarba-and-Picano-JT-2020,Berghout-et-al-JFM-2019}, numerical methods for two-phase \cite{Scapin-et-al-JCP-2020,Liu-et-al-JCP-2021} or non-Newtonian flows \cite{Ahmed-et-al-CF-2020}.\par
In the near future, and in line with recent efforts in the fluid dynamics community, \emph{SNaC} will be ported for massively parallel calculations on many Graphics Processing Units (GPUs) \cite{Zhu-et-al-CPC-2018,Costa-et-al-CAMWA-2021,Bernardini-et-al-CPC-2021,Ha-et-al-CPC-2021}. In addition to this major milestone, an implementation of the multigrid solver will be sought which directly solves a batch of small linear systems, so that the inhomogeneity of the reduced 2D linear systems is fully covered without compromising the parallel performance.
\section*{Acknowledgments}
I would like to thank Luca Brandt for interesting discussions, and the first users of \emph{SNaC} from KTH Mechanics, Arash Banaei, Nazario Mastroianni, and Nicol\`o Scapin for the invaluable feedback and testing. Dr.~Rob Falgout from Lawrence Livermore National Laboratory is thanked for suggesting the ``sliced pencils'' approach using \emph{hypre} in Algorithm~\ref{alg:fast_poi_solver_par}, as an alternative to Algorithm~\ref{alg:fast_poi_solver_par_oth}. Prof.~Fernando Pinho from University of Porto (FEUP) is thanked for kindly providing the validation data from Ref.~\cite{Miranda-et-al-IJNMF-2008}. Finally, the two anonymous reviewers are thanked for the useful feedback on an earlier version of this manuscript. The computing time for the scaling tests was provided by the Swedish National Infrastructure for Computing (SNIC), and the National Infrastructure for High-Performance Computing and Data Storage in Norway, (Sigma2). This work was supported by the University of Iceland Recruitment Fund grant No.~1515-151341, \emph{TURBBLY}.
\appendix
\section{Alternative approach for solving the Poisson equation}\label{sec:poi_other}
\setcounter{figure}{0}
Here we present an alternative approach that may be employed for the solution of $n$, 2D linear systems (Eq.~\ref{eqn:poi_reduced}), using a slab domain decomposition. Unlike Algorithm~\ref{alg:fast_poi_solver_par}, where an appropriate value of $p$ needs to be determined, this approach does not require tuning. Let $\underline{n}^b = \left[n_1^b,n_2,n_3^b\right]$ be the number of grid points in each direction specific to block $b$, with the same number of grid points in the synthesis direction (here taken again as $x_2$, so $n_2^b=n_2$). Instead of solving the 2D linear systems sequentially in a pencil domain decomposition, we follow the FFT-based synthesis by a \emph{redistribution} of the domain decomposition within each block to a \emph{slab-like} configuration, as illustrated in the right drawing of Fig.~\ref{fig:transpose_to_slab}.\par
In this configuration, each subdomain $m$ has a size $\left[n_1^b,n_2^{m},n_3^b\right]$, i.e., with the points along $x_2$ decomposed by the total number of tasks within block $b$. This operation is employed using an \emph{all-to-all}\footnote{In practice, implemented using \texttt{MPI{\us}Alltoallw} and subarray MPI derived types.} collective operation within the group of tasks of each block (i.e., under \verb|comm_block|). Solving the iterative system using this configuration has clear advantages: first, the communication required for each 2D system is much smaller; second, the solution of the different $n_2$ systems is now \emph{parallel}, in batches of size $n_2^{m}$. We will see that these advantages justify the overhead of the \emph{all-to-all} collective, especially if $n_2$ is large enough. Besides the collective operations, a downside of this approach is the hard limit of the number of tasks per block, which cannot exceed $n_2$ in this example. However, this restriction can be significantly relaxed by leveraging shared-memory parallelization.
\begin{figure}[hbt!]
  \centering
  \begin{tikzpicture}
    \node[draw=none,fill=none] at (0,0) {\includegraphics[width=0.95\columnwidth]{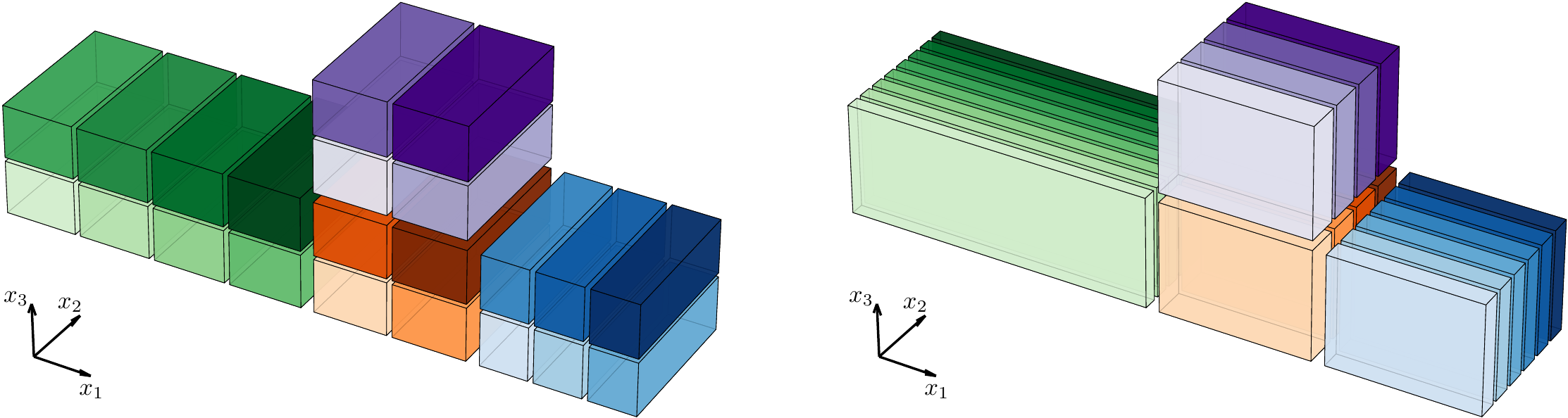}};
    \draw[to-to,thick] (-2.0,2.0) to [in=160,out=20] ( 2.0,2.0);
    \node at (0.,2.1) {\scriptsize \emph{all-to-all} collective};
    \node at (0.,1.8) {\tiny (within \verb|comm_block|)};
    \node at (-4.5,-2.2) {\scriptsize grid cells for \emph{pencil} subdomain $m$: $\left[n_1^m\times n_2\times n_3^m\right]$};
    \node at (-4.5,-2.6) {\scriptsize large decomposition $\to$ $n_2^b\gg n_1^m\,\mathrm{and}\,n_3^m $~~~~~~~~~~~~~~~~~};
    \node at ( 4.2,-2.2) {\scriptsize grid cells for \emph{slab} subdomain $m$: $\left[n_1^b\times n_2^m\times n_3^b\right]$};
    \node at ( 4.2,-2.6) {\scriptsize large decomposition $\to$ $n_1^b\,\mathrm{and}\,n_3^b \gg n_2^m$~~~~~~~~~~~~~~};
  \end{tikzpicture}
  \caption{Illustration of two different domain decompositions which may be used to solve the FFT-accelerated Poisson equation, assuming $x_2$ as the FFT synthesis direction. Different colors distinguish the tasks in different blocks, while different lightness marks the MPI tasks within a block. The left side shows a pencil decomposition, required for the FFT-based synthesis, and the right side shows a slab decomposition within each block used for the solution of Eq.~\eqref{eqn:poi_reduced}. The redistribution from one decomposition to the other requires a collective \emph{all-to-all} operation, performed within the group of ranks in each block (\texttt{comm\_block}).}\label{fig:transpose_to_slab}
\end{figure}
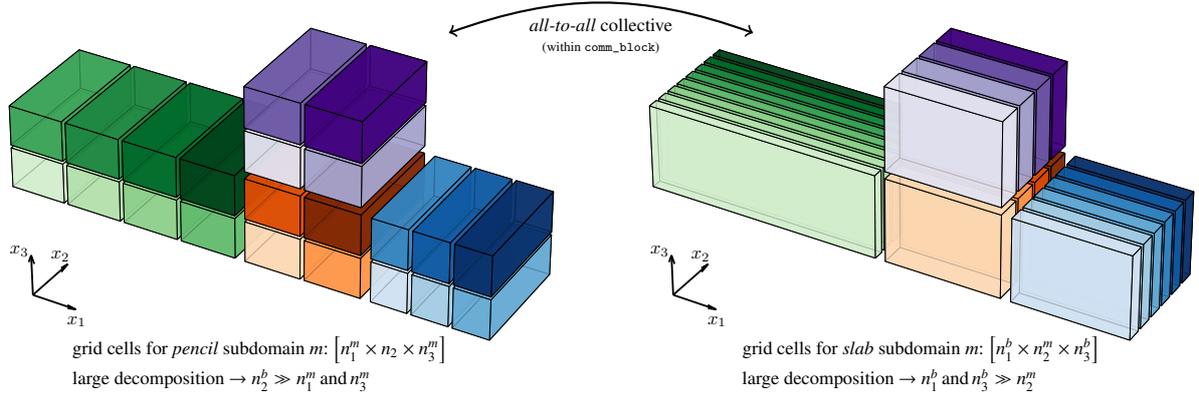\par
As Fig.~\ref{fig:transpose_to_slab} illustrates, the slab decomposition is not required to be congruent among the different blocks -- domains with larger values of $n_1^b\times n_3^b$ can be more decomposed, to ensure load balancing. This means that the communicator associated with the iterative solution of the $n_2^m$ 2D systems to be passed to \emph{hypre} cannot be \verb|comm_block|. Instead, an array of MPI communicators \verb|comm_slab(:)| is determined, where each element encapsulates the tasks in charge of the 2D linear system associated with the plane with index $j$. The overall approach for the parallel FFT-based solution of the Poisson equation is presented in Algorithm~\ref{alg:fast_poi_solver_par_oth}.\par
\begin{algorithm}[hbt!]
  \caption{Sequence of operations performed per task for the parallel solution of the Poisson equation (Eq.~\eqref{eqn:poi}) with FFT-based synthesis, using a pencil$\leftrightarrow$slab data redistribution.}\label{alg:fast_poi_solver_par_oth}
  \begin{algorithmic}
    \setstretch{1.0}
    \small
    \State \textbf{do} {$i=1$ to $n_1^m$ and $k=1$ to $n_3^m$}
    \State \quad forward FFT-based transform along $x_2$ of right-hand-side of Eq.~\eqref{eqn:poi}: $\hat{f}_{i,1 \dots n_2,k} = \mathcal{F}_{x_2}(f_{i,1 \dots n_2,k})$
    \State \textbf{end do}
    \State \emph{pencil}$\to$\emph{slab} redistribution within each block (i.e., within \verb|comm_block|) to obtain $\hat{f}_{1 \dots n_1^b,1\dots n_2^m,1\dots n_3^b}$
    \State \textbf{do} $j=1$ to $n_2^m$
    \State \quad solve Eq.~\eqref{eqn:poi_reduced} within \verb|comm_slab(j)| using a geometric multigrid solver to obtain $\hat{\Phi}_{1 \dots n_1^b,j,1 \dots n_3^b}$
    \State \textbf{end do}
    \State \emph{slab}$\to$\emph{pencil} redistribution within each block (i.e., within \verb|comm_block|) to obtain $\hat{\Phi}_{1 \dots n_1^m,1\dots n_2,1\dots n_3^m}$
    \State \textbf{do} $i=1$ to $n_1^m$ and $k=1$ to $n_3^m$
    \State \quad backward FFT-based transform along $x_2$ of the solution: $\Phi_{i,1 \dots n_2,k} = \mathcal{F}_{x_2}^{-1}(\hat{\Phi}_{i,1 \dots n_2,k})$
    \State \textbf{end do}
  \end{algorithmic}
\end{algorithm}
\section{On the performance effects of the pencil slicing parameter $p$}\label{sec:pencil_slicing_analysis}
\setcounter{figure}{0}
Here we analyze the effect of the number of pencil slices, $p$, in the performance of Algorithm~\ref{alg:fast_poi_solver_par}. Recall that, after performing the FFT-based synthesis of the Poisson problem, $n_l$ 2D independent linear systems are to be solved using an iterative (multigrid) method. The main diagonal of each system varies along the synthesis direction $x_l$, according to the eigenvalue $\lambda$ (recall Eq.~\eqref{eqn:poi_reduced}), resulting in a varying ``diagonal dominance'' of the problems along $x_l$. Hence, for the same iterative error tolerance, the number of iterations will vary among 2D problems, being larger the less ``diagonally dominant'' the problem is. This is illustrated in Fig.~\ref{fig:pencil_slicing_analysis}(a) for one of the cases addressed in \S\ref{sec:results} -- after Fourier synthesis, the higher the wavenumber $\kappa$, the larger the magnitude of $\lambda$, and the lower the required number of iterations $N_{iter}$ to solve the reduced Helmholtz problems.\par
The sliced pencils approach (Algorithm~\ref{alg:fast_poi_solver_par}) aims at covering this problem inhomogeneity, while ensuring a significant load per task. If $p = n_l$, this inhomogeneity is perfectly covered by solving the problem plane-by-plane, but the load per task is too small, and communication overwhelms computation\footnote{We recall that this corresponds to the naive approach in \S\ref{sec:mg_w_fft}, except that here a planar problem is still treated as a 3D problem decoupled along $x_l$.}. Conversely, the limit of $p=1$ -- a single, large 3D problem -- results in a lot of unnecessary work, since the number of iterations will be dictated by the slowest-converging 2D problem. Hence, there is an optimal value of $p$ which shows good compromise in terms of load per task and capturing the problem inhomogeneity. Fig.~\ref{fig:pencil_slicing_analysis}(b) shows the influence of this parameter in the wall-clock time per time step of one of the problems addressed above (see the figure caption), where $p\approx 16$ seems to show a good compromise. A possible improvement in the present method is performing an autotuning step at the beginning of the calculation, which optimally distributes the partitioning to cover inhomogeneous distribution of $N_{iter}$, possibly unevenly.
\begin{figure}[hbt!]
  \centering
  \includegraphics[width=0.49\columnwidth]{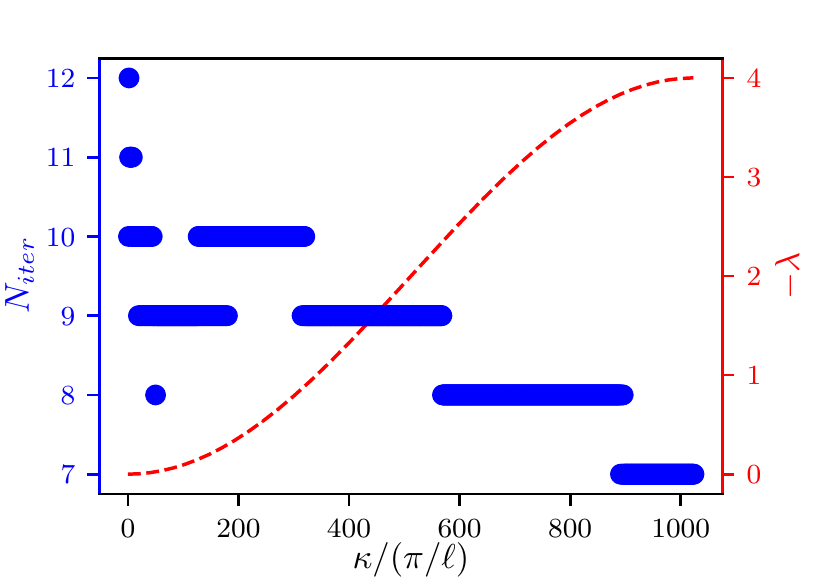}\hfill
  \includegraphics[width=0.49\columnwidth]{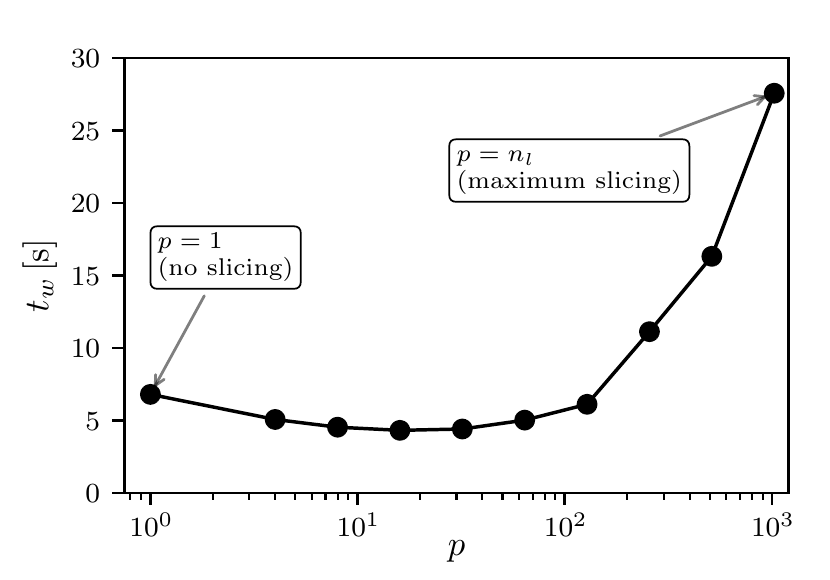}
  \caption{Lid-driven cavity flow setup with $N=1024^3$ points, solved with FFT-based synthesis along $x$. (a): evolution of the number of iterations in the PFMG solver, $N_{iter}$, required to solve each 2D problem after FFT-based synthesis, for a fixed iterative error tolerance of $10^{-4}$ (blue), and the corresponding eigenvalue $\lambda$ associated with the cosine series expansion along the $x$ direction (red), both as a function of the wavenumber $\kappa$ (recall Table~\ref{tbl:operators}). (b): wall-clock time as a function of the pencil partitioning parameter $p$ when solving the problem with the sliced pencils approach in Algorithm~\ref{alg:fast_poi_solver_par} with $N_{CPU}=1024$.}\label{fig:pencil_slicing_analysis}
\end{figure}
\section{Performance assessment at extreme scales}\label{sec:scaling_big}
\setcounter{figure}{0}
This section studies the weak and strong scaling performance of the overall implementation at more extreme scales. While we restrict ourselves to the lid-driven cavity flow case for simplicity, we expect it to be representative of other multi-block configurations, as long as the data is evenly distributed among tasks.\par
\begin{figure}[hbt!]
  \centering
  \includegraphics[width=0.49\columnwidth]{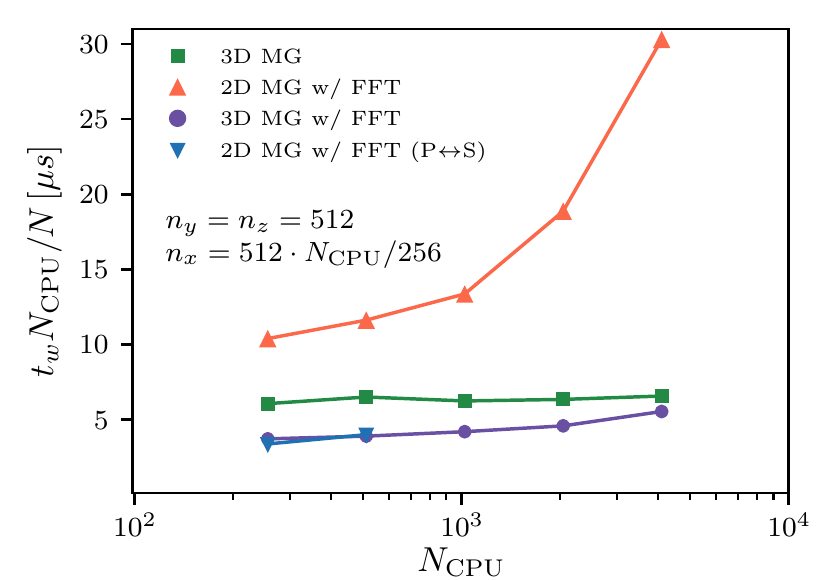}\hfill
  \includegraphics[width=0.49\columnwidth]{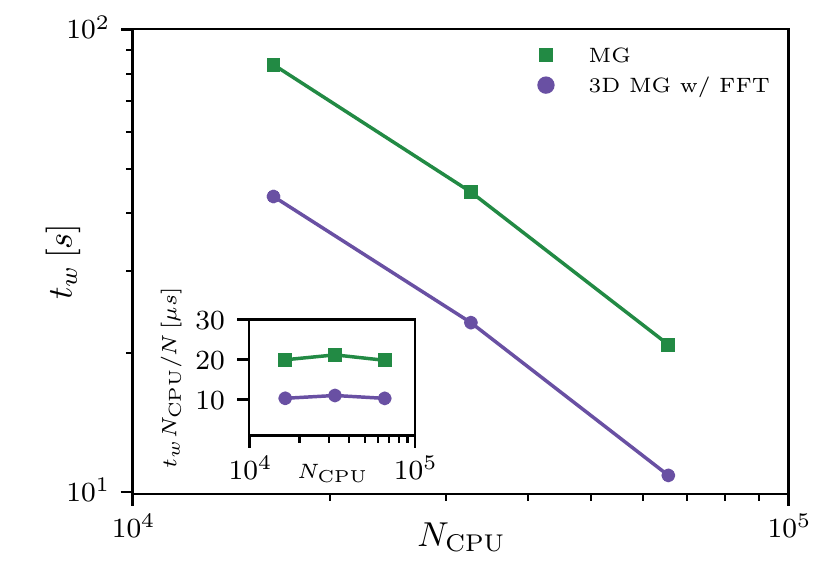}
  \caption{(a): weak scaling of the numerical algorithm up to $8192$ cores and $4.3\cdot10^9$ grid points. The domain length and number of grid points were successively extended along the $x$ direction, so as to maintain the grid spacing and number of grid points per task. $t_{w}$ denotes wall-clock time in microseconds per grid point, per time step, and $N_{CPU}$ the number of cores. Here horizontal lines correspond to ideal scaling. (b): strong scaling of the numerical algorithm up to $65\,536$ cores, for a lid-driven cavity setup with $4096^3$ grid cells. $t_{w}$ denotes wall-clock time in seconds/time step/task (i.e., three Runge-Kutta substeps), and $N_{CPU}$ the number of cores. The inset shows, again, the wall-clock time per grid point, per time step in microseconds, where horizontal lines mean ideal scaling. Both cases pertain to a lid-driven cavity flow with $x$-oriented pencil subdomains. Legend notation -- see the beginning of \S\ref{sec:performance}.}\label{fig:scaling_extreme}
\end{figure}

Fig.~\ref{fig:scaling_extreme}(a) presents the weak scaling of the same problem as Fig.~\ref{fig:scaling_ldc}, starting from a $512^3$ grid, with both the grid spacing and number of points per task fixed. Expectedly, while the scaling is poor for the naive \yofft case, the other cases show good performance. This suggests that the strong scaling performance shown in Fig.\ref{fig:scaling_ldc} should still hold for larger problem sizes and number of CPUs.\par
To confirm this, we tested the strong scaling overall algorithm at extreme scales, on a $4096^3$ box in up to $65\,536$ CPUs. The simulations were carried out on the Betzy supercomputer, based in Norway (Bull Sequana XH2000, AMD EPYC 7742 64C 2.25GHz, Mellanox HDR Infiniband), and the results are shown in Fig.~\ref{fig:scaling_extreme}(b). The \nofft and the preferred FFT-accelerated case, \yrfft, are the only ones who were able to run efficiently (or at all) at these scales. Both cases show very good scaling performance, as highlighted by the figure inset. Moreover, despite the different hardware, the wall-clock time per grid point is within the same order-of-magnitude as that of the weak scaling plot, with slightly larger values which are expected, since the finer grid requires more iterations in the Poisson solver. These observations are somewhat consistent with the excellent performance of the \emph{hypre} library at extreme scales (see, e.g., \cite{Baker-et-al-2012}) which, as we have shown, holds the largest share of compute time within the calculation timeline.\par
Not surprisingly, the naive \yofft implementation performed poorly at these scales, with the runs either failing or costing no less than an order of magnitude more computing time than the case without FFT acceleration (not shown). On the other hand, the FFT-accelerated slab-decomposed case (\yefft; Algorithm~\ref{alg:fast_poi_solver_par_oth}) failed to run due to the overflow of the integer displacement vectors in the \texttt{MPI{\us}Alltoallw} collective, which is a known limitation of the MPI library; see \cite{Hammond-et-al-IEEE-2014}. This issue also affected the weak scaling plot in panel (a) of Fig.~\ref{fig:scaling_extreme}. Fortunately, large counts are supported in the latest MPI Standard $4.0$ \cite{mpi40}, which will resolve this issue without major changes in the current implementation.
%
\bibliography{bibfile}
\end{document}